\documentclass[final]{cvpr}

\usepackage{times}
\usepackage[caption=false]{subfig}

\usepackage{epsfig}
\usepackage{graphicx}
\usepackage{amsmath}
\usepackage{amssymb}
\usepackage{multirow}
\usepackage{makecell}
\usepackage{pifont}
\usepackage{color}
\usepackage[abs]{overpic}

\definecolor{citegreen}{RGB}{50,175,50}
\usepackage[pagebackref,
            breaklinks,
            colorlinks,
            citecolor=citegreen]{hyperref}

\def\cvprPaperID{0335} 
\def\confName{CVPR}
\def\confYear{2022}

\graphicspath{{./figs/}{./figs/result/}}

\newcommand{\cmark}{\ding{51}}%
\newcommand{\xmark}{\text{\ding{55}}}
\newcommand{\subtitle}[1]{\noindent \textbf{#1}}
\newcommand{\methodname}{E$^2$FGVI}
\newcommand*\samethanks[1][\value{footnote}]{\footnotemark[#1]}

\begin{document}

\title{Towards An End-to-End Framework for Flow-Guided Video Inpainting}

\author{
  Zhen Li\textsuperscript{1}\thanks{Equal contribution} \quad
  Cheng-Ze Lu\textsuperscript{1}\samethanks \quad
  Jianhua Qin\textsuperscript{2} \quad
  Chun-Le Guo\textsuperscript{1}\thanks{C.L. Guo is the corresponding author.} \quad
  Ming-Ming Cheng\textsuperscript{1} \\
  \textsuperscript{1}TMCC, CS, Nankai University  \quad
  \textsuperscript{2}Hisilicon Technologies Co. Ltd. \\
  {\tt\small
    zhenli1031@gmail.com,
    czlu919@outlook.com,
    qinjianhua@hisilicon.com
  } \\
  {\tt\small
  \{guochunle, cmm\}@nankai.edu.cn
  }
}

\maketitle

\begin{abstract}
    Optical flow, which captures motion information across frames, is exploited in recent video inpainting methods through propagating pixels along its trajectories. 
    However, the hand-crafted flow-based processes in these methods are applied separately to form the whole inpainting pipeline.
    Thus, these methods are less efficient and rely heavily on the intermediate results from earlier stages.
    In this paper, we propose an End-to-End framework for Flow-Guided Video Inpainting (\methodname) through elaborately designed three trainable modules, 
    namely, flow completion, feature propagation, and content hallucination modules.
    The three modules correspond with the three stages of previous flow-based methods but can be jointly optimized, leading to a more efficient and effective inpainting process.
    Experimental results demonstrate that the proposed method outperforms state-of-the-art methods both qualitatively and quantitatively and shows promising efficiency. 
    The code is available at \url{https://github.com/MCG-NKU/E2FGVI}.
    \vspace{-5mm}
\end{abstract}


\section{Introduction}
Video inpainting aims to fill up the ``corrupted'' regions with plausible and coherent content throughout video clips.
It is widely applied to real-world applications such as object removal~\cite{7112116}, video restoration~\cite{lee2019cpnet}, and video completion~\cite{chang2019free,oh2019onion}.
Despite the significant progress made in image inpainting~\cite{yu2018generative,yu2018free,pathak2016context}, video inpainting remains full of challenges due to the complex video scenarios and deteriorated video frames. 
Directly performing image inpainting on each frame independently tends to generate temporally inconsistent videos and results in severe artifacts.
Both spatial structure and temporal coherence are required to be considered in high-quality video inpainting.
Recent progress in deep learning motivates researchers to exploit more effective solutions~\cite{newson2014video,wang2019video,kim2019deep,chang2019learnable,chang2019free,lee2019cpnet,Xu_2019_CVPR,yan2020sttn,Liu_2021_FuseFormer,Gao-ECCV-FGVC}.

\begin{figure}[t]
    \centering
    \subfloat[]{\includegraphics[width=0.47\textwidth]{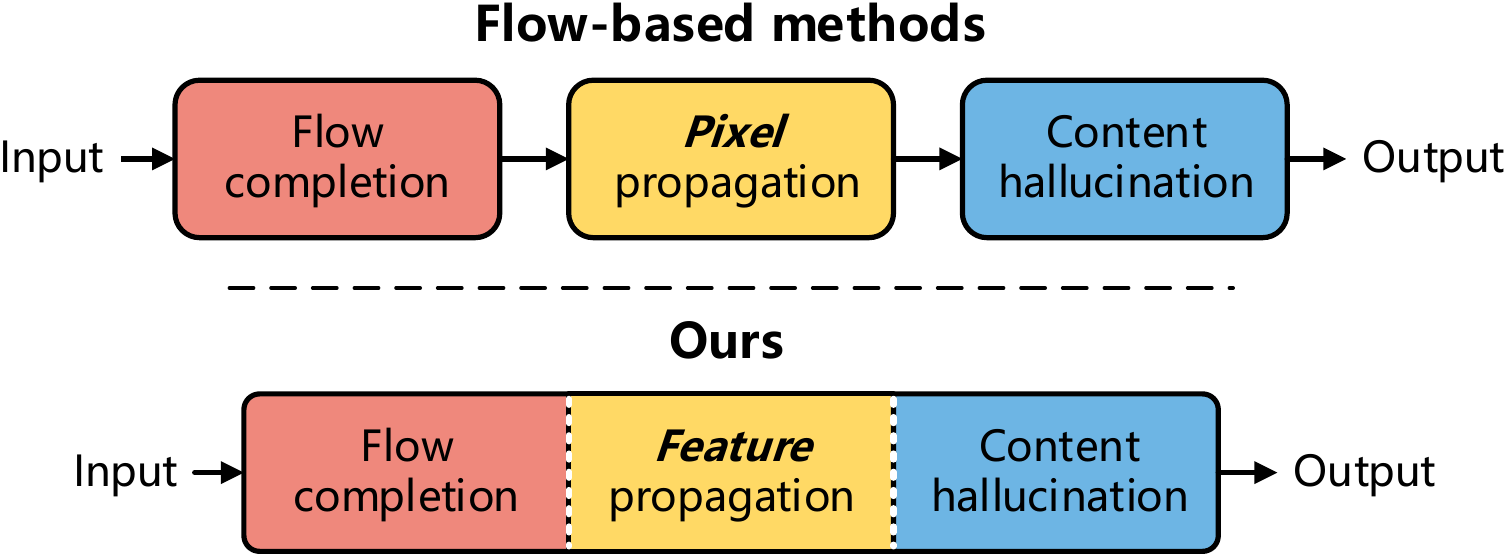}}\\
    \vspace{-3mm}
    \subfloat[]{\includegraphics[width=0.47\textwidth]{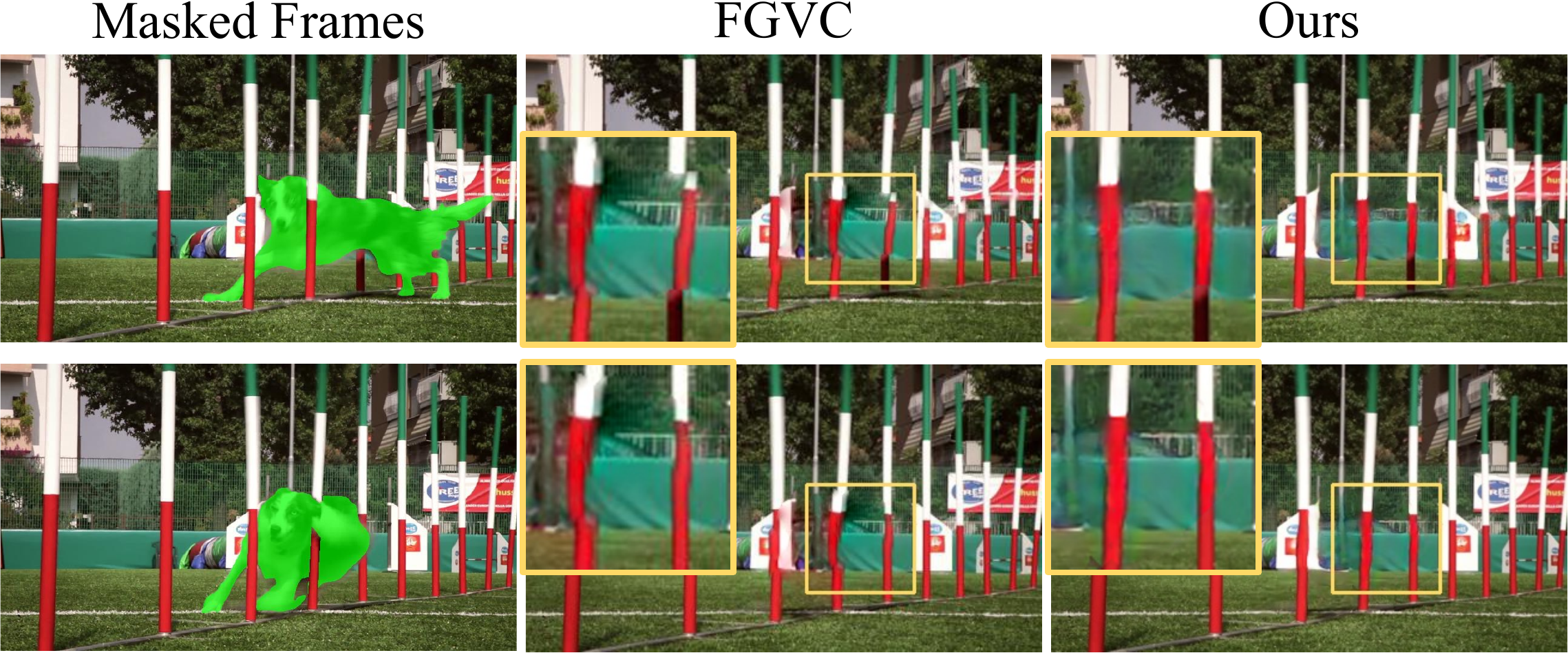}}
    \caption{
        (a) The general pipelines of flow-based methods~\cite{Xu_2019_CVPR,Gao-ECCV-FGVC} and ours.
        While previous flow-based methods conduct the three stages separately, our corresponding modules work in an end-to-end manner.
        (b) A qualitative comparison of our approach with a state-of-the-art flow-based method FGVC~\cite{Gao-ECCV-FGVC}.
        Due to the error accumulation and ignoring temporal information during content hallucination, FGVC fails to generate faithful and temporally consistent results compared with our method.
        }
    \label{fig:intro}
    \vspace{-4mm}
\end{figure}

Among them, typical flow-based methods~\cite{Xu_2019_CVPR,Gao-ECCV-FGVC} consider video inpainting as a pixel propagation problem to naturally preserve the temporal coherence.
As shown in \figref{fig:intro}~(a), these methods can be decomposed into three inter-related stages.
(1) \textit{Flow completion}: The estimated optical flow needs to be completed first because the absence of flow fields in corrupted regions will influence the latter processes.
(2) \textit{Pixel propagation}: They fill the holes in corrupted videos by bidirectionally propagating pixels in the visible areas with the guidance of the completed optical flow.
(3) \textit{Content hallucination}: After propagation, the remaining missing regions can be hallucinated by a pre-trained image inpainting network~\cite{yu2018generative,yu2018free}.

Unfortunately, even though impressive results can be obtained, the whole flow-based inpainting process must be carried out separately as many hand-crafted operations 
(\eg, Poisson blending, solving sparse linear equations, and indexing per-pixel flow trajectories) are involved in the first two stages.
The isolated processes raise two main problems.
One is that the errors that occur at earlier stages would be accumulated and amplified at subsequent stages, which further influences the final performance significantly.
Specifically, the inaccurate flow estimation would mislead the propagation of pixels and further confuse the stage of content hallucination, producing unfaithful inpainting results.
Second, these complex hand-designed operations only can be processed without GPU acceleration.
The whole procedure of inferring video sequences, therefore, is very time-consuming.
Taking DFVI~\cite{Xu_2019_CVPR} as an example, completing one video with the size of 432 $\times$ 240 from DAVIS~\cite{perazzi2016benchmark}, 
which contains about 70 frames, needs about 4 minutes\footnote{We test it on Intel(R) Core(TM) i7-6700K CPU with a single NVIDIA Titan Xp GPU.}, which is unacceptable in most real-world applications.
Besides, except for the above-mentioned drawbacks, only using a pretrained image inpainting network at the content hallucination stage ignores the content relationships across temporal neighbors, leading to inconsistent generated content in videos (see \figref{fig:intro}~(b)).

To address these flaws, in this paper, we carefully design three trainable modules, including (1) flow completion, (2) feature propagation, and (3) content hallucination modules which simulate corresponding stages in flow-based methods and further constitute an End-to-End framework for Flow-Guided Video Inpainting (\methodname).
Such close collaboration between the three modules alleviates the excessive dependence of intermediate results in the previously independently developed system~\cite{kim2019deep,Xu_2019_CVPR, Gao-ECCV-FGVC, zou2020progressive, lao2021flow} and works in a more efficient manner.

To be specific, for the \textit{flow completion} module, we directly employ it on the masked videos for one-step completion instead of multiple complex steps.
For the \textit{feature propagation} module, in contrast to the pixel-level propagation, our flow-guided propagation process is conducted in the feature space with the assistance of deformable convolution.
With more learnable sampling offsets and feature-level operations, the propagation module releases the pressure of inaccurate flow estimation.
For the \textit{content hallucination} module, we propose a temporal focal transformer to effectively model long-range dependencies on both spatial and temporal dimensions.
Both local and non-local temporal neighbors are considered in this module, leading to more temporally coherent inpainting results.

Experimental results demonstrate that 
our framework enjoys the following two strengthens:
\vspace{-2mm}
\begin{itemize}
    \setlength{\itemsep}{-2pt}
    \setlength{\parsep}{-3pt}
    \item State-of-the-art accuracy: 
    Taking comparisons with previous state-of-the-art (SOTA) methods, the proposed \methodname ~achieves significant improvements on two common distortion-oriented metrics (\ie, PSNR and SSIM~\cite{wang2004image}), one popular perception-oriented index (\ie, VFID~\cite{NEURIPS2018_d86ea612}), and one temporal consistency measurement (\ie, $E_{warp}$~\cite{Lai-ECCV-2018}).
    \item High efficiency: 
    Our method processes $432\times240$ videos at 0.12 seconds per frame on a Titan Xp GPU, which is nearly $15\times$ faster than previous flow-based methods.
    In contrast to methods that also can be end-to-end deployed, our method shows comparable inference time.
    Besides, our method has the lowest computational complexity (FLOPs) among all compared SOTA methods.
\end{itemize}
\vspace{-2mm}
We hope the proposed end-to-end framework with the aforementioned advantages could serve as a strong baseline for the video inpainting community.
\begin{figure*}[t]
    \centering
    \begin{overpic}[width=\textwidth]{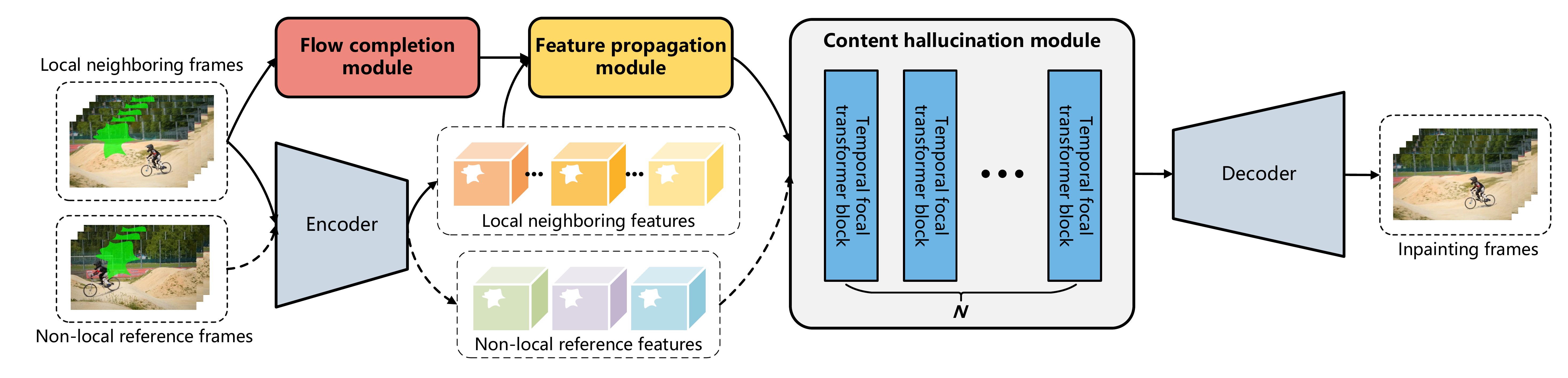}
    \end{overpic}
    \caption{Overview of the proposed End-to-End framework for Flow-Guided Video Inpainting (\methodname).
     It consists of 1) a frame-level content encoder, 2) a flow completion module, 3) a feature propagation module, 4) a content hallucination module which is composed of multiple temporal focal transformer blocks, and 5) a frame-level decoder.}
    \label{fig:pipeline}
    \vspace{-4mm}
\end{figure*}

\section{Related Work}
\subtitle{Video inpainting.}
Building upon the development of deep learning, great progress has been made in video inpainting.
These methods can be roughly divided into three classes: 3D convolution-based~\cite{wang2019video,chang2019learnable,hu2020proposal}, flow-based~\cite{Xu_2019_CVPR,Gao-ECCV-FGVC}, and attention-based methods~\cite{lee2019cpnet, li2020short, Liu_2021_FuseFormer, yan2020sttn}.
Some methods~\cite{chang2019free,kim2019deep,wang2019video,lee2019cpnet} employing 3D convolution and attention usually yield temporally inconsistent results due to the limited temporal receptive fields.
To generate more temporal coherence results, many works~\cite{kim2019deep,zou2020progressive} regard optical flows as strong priors for video inpainting and incorporate them into the network.
However, directly computing optical flows between images within invalid regions is extremely difficult as these regions themselves become occlusion factors, restricting the performance.
Recent flow-based methods~\cite{Xu_2019_CVPR, Gao-ECCV-FGVC} perform flow completion first and use the completed optical flows to propagate indexed pixels along their trajectories.
Instead of conducting hand-crafted pixel-level propagation, we design an end-to-end trainable framework that performs the propagation process at the feature space.
Besides, our method benefits from recent advances in using transformers to improve the inpainting results~\cite{yan2020sttn, liu2021decoupled, Liu_2021_FuseFormer}.

\subtitle{Flow-based video processing.}
The motion information across frames well assists many video-related tasks, such as video understanding~\cite{carreira2017quo, lin2019tsm}, video segmentation~\cite{tsai2016video, cheng2017segflow}, video object detection~\cite{zhu2017flow},
depth estimation~\cite{godard2019digging, luo2020consistent}, video super-resolution~\cite{xue2019video,chan2021basicvsr}, frame interpolation~\cite{jiang2018super,lee2020adacof}, \etc
Specifically, many video restoration and enhancement algorithms~\cite{kim2018spatio,xue2019video,tian2020tdan,pan2020cascaded,chan2021basicvsr} rely on optical flow to perform alignment for compensating the information between frames.
Recent works~\cite{wang2019edvr, xiang2020zooming, lee2020adacof, xu2021temporal, chan2021basicvsr} leverage deformable convolution~\cite{zhu2019deformable} to simulate the behavior of optical flow but with more learnable offsets for more effective alignment.
Our works also share the same merit as these works.

\subtitle{Vision transformer.}
Recently, Transformer~\cite{NIPS2017_3f5ee243} has gained much attention in the vision community.
Vision Transformer~\cite{dosovitskiy2021an} and its follow-ups~\cite{pmlr-v139-touvron21a,yuan2021tokens,liu2021Swin,yang2021focal, guo2022visual}
achieve an impressive performance on image and video representation learning~\cite{chen2020generative,desai2021virtex,liu2021video,patrick2021keeping}, 
image generation~\cite{parmar2018image}, object detection~\cite{carion2020end,zhu2020deformable}, and many other applications~\cite{choromanski2020rethinking,cheng2021maskformer,liang2021swinir,guo2022attention}.
Because of the quadratic complexity of self-attention, many works deployed effective window-based attentions~\cite{liu2021Swin,yang2021focal,dong2021cswin} to reduce its computational complexity while improving the model's capability with the limited receptive fields.
Swin Transformer~\cite{liu2021Swin} strengthens local connections by computing self-attention through shifting local windows.
Focal Transformer~\cite{yang2021focal} introduces focal self-attention, which enhances the global-local interactions.
\section{Method}
Given a corrupted video sequence $\{X^{t} \in \mathbb{R}^{ H\times W\times 3} \mid t=1 \dots T\}$ with sequence length $T$ and corresponding frame-wise binary masks $\{M^{t} \in \mathbb{R}^{ H\times W\times 1} \mid t=1 \dots T\}$,
we aim at synthesizing faithful content which is consistent in both space and time dimensions within the corrupted (masked) areas.
In the following, we discuss the main components of our method.
First, we use a context encoder, which encodes all corrupted frames into lower-resolution features for computational efficiency at subsequent processing.
Second, we extract and complete the optical flow between local neighbors through a flow completion module (Sec.~\ref{sec:flow_alignment}).
Third, the completed optical flow assists the features extracted from local neighbors to accomplish feature alignment and bidirectional propagation (Sec.~\ref{sec:flow_alignment}).
Fourth, multi-layer temporal focal transformers perform content hallucination by combining propagated local neighboring features with non-local reference features. (Sec.~\ref{sec:focal_transformer}). 
Finally, a decoder up-scales the filled features and reconstructs them to a final video sequence $\{\hat{Y}^{t} \in \mathbb{R}^{ H\times W\times 3} \mid t=1 \dots T\}$.

\figref{fig:pipeline} shows the whole pipeline of the proposed \methodname. 
It is worth noticing that all modules are differentiable and constitute an end-to-end trainable architecture.

\subsection{Flow completion and feature propagation}
\label{sec:flow_alignment}
In this section, we will detail the proposed flow-related operations.
Note that we only apply flow-based modules on the features extracted from local neighboring frames because the flow estimation is substantially degraded or even fails because of the presence of large motion, which frequently occurs in non-local frames.
Besides, the flow-related operations are given at lower-resolution space for computational efficiency.

\subtitle{End-to-end flow completion.}
Before flow prediction, we first downsample the original corrupted frames $X^{t}$ at 1/4 resolution, which matches the spatial resolution of encoded low-resolution features.
The downsampled frames are denoted as $X^{t}_{\downarrow} \in \mathbb{R}^{ \frac{H}{4} \times \frac{W}{4}\times 3}$.
The flow prediction between adjacent frames $i$ and $j$ is computed by a flow estimation network $\mathcal{F}$:
\begin{equation}
    \small
    \hat{F}_{i\rightarrow j} = \mathcal{F}(X^{i}_{\downarrow}, X^{j}_{\downarrow}).
    \label{eq:flow}
\end{equation}
We initialize the network using pretrained weights from a lightweight flow estimation network to resort to its rich knowledge about optical flows.

Following most flow-based video inpainting methods~\cite{Gao-ECCV-FGVC,Xu_2019_CVPR}, we estimate both forward flow $\hat{F}_{t\rightarrow t+1}$ and backward flow $\hat{F}_{t\rightarrow t-1}$ through Eq.~\eqref{eq:flow} for flow-guided bidirectional propagation. 
Since the missing areas in corrupted videos become occlusion factors for flow estimation, which severely affects the quality of estimated flow, we need to restore the forward and backward flow before using them for feature propagation.
For simplicity, we use L1 loss\footnote{Other loss functions can also be used in Eq.~\eqref{eq:loss_flow}, but we do not observe significant improvements on the final inpainting performance.} to restore the bidirectional flows:
\begin{equation}
    \label{eq:loss_flow}
    \small
    \mathcal{L}_{flow}=\sum_{t=1}^{T-1} \|\hat{F}_{t\rightarrow t+1}-F_{t\rightarrow t+1}\|_{1} + \sum_{t=2}^{T} \|\hat{F}_{t\rightarrow t-1}-F_{t\rightarrow t-1}\|_{1},
\end{equation}
where $F_{t\rightarrow t+1}$ and $F_{t\rightarrow t-1}$ are the ground truth forward and backward flow, respectively, which are calculated from original uncorrupted videos.

Our flow completion module differs from DFVI~\cite{Xu_2019_CVPR} and FGVC~\cite{Gao-ECCV-FGVC} from two main aspects.
(1) DFVI and FGVC deploy the flow completion network and propagation algorithm separately. 
In contrast, our flow completion module can be trained with other network components in an end-to-end manner, which facilitates the module to generate task-oriented flows~\cite{xue2019video}.
(2) The flow completion in DFVI and FGVC is less efficient ($>$ 0.4s/flow) because they need to initialize the flow first and then refine the initialized flow with multiple stages,
while we estimate and complete the flow in only one feed-forward pass with much faster speed ($<$ 0.01s/flow).

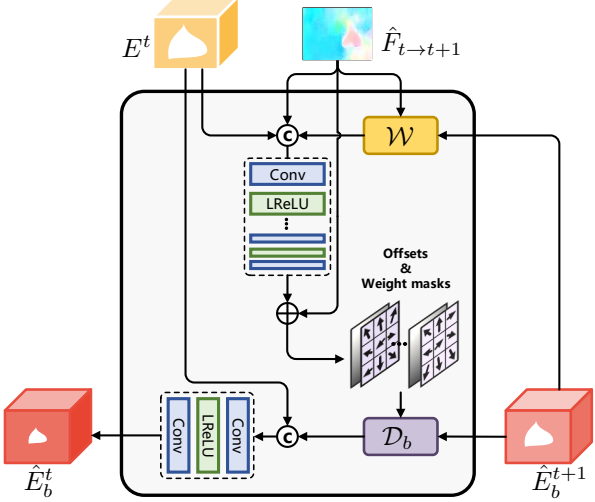
\begin{figure}[t]
    \centering
    \begin{overpic}[width=0.45\textwidth]{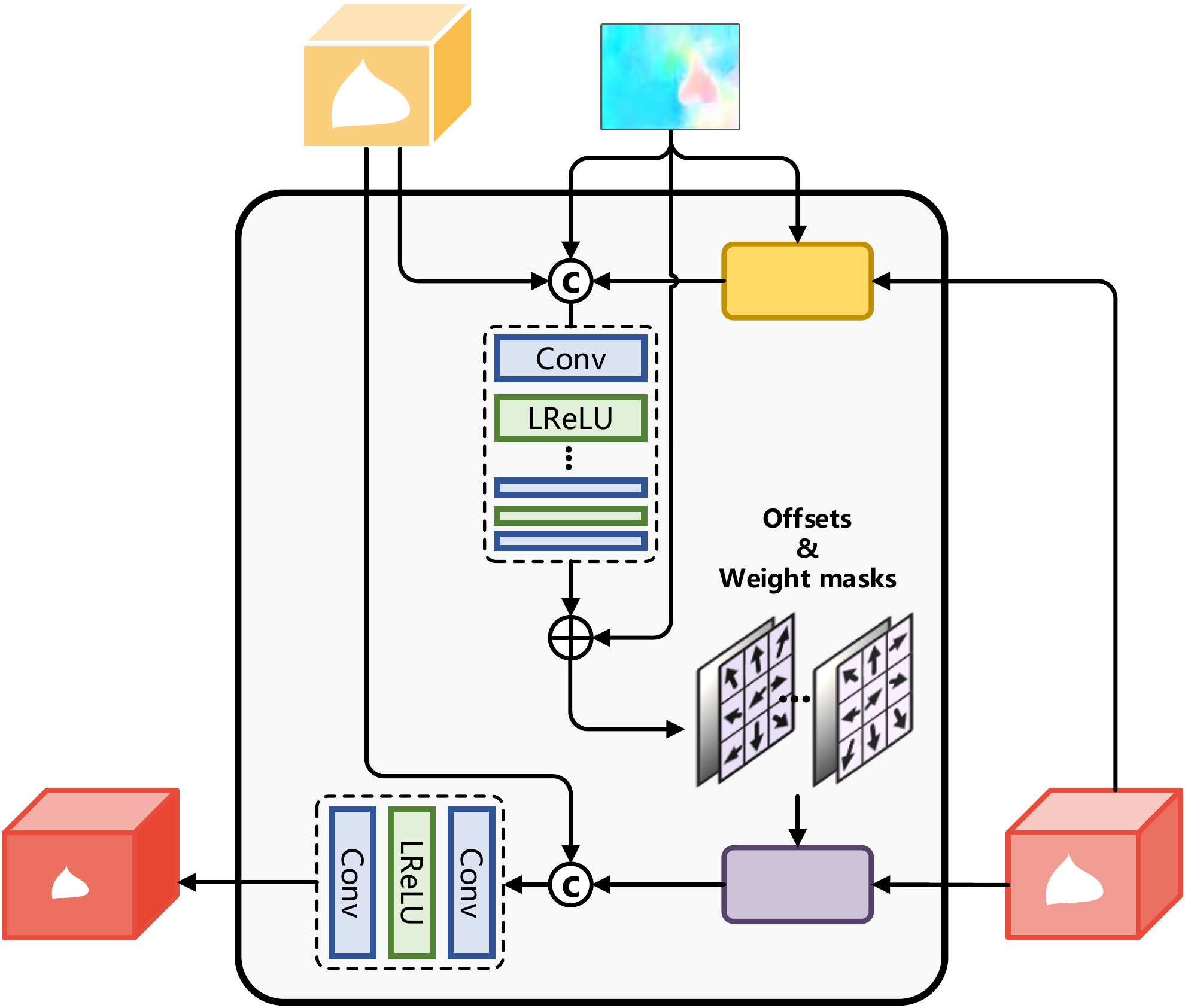}
        \put(143.5,20){\color{black}{$\mathcal{D}_{b}$}}
        \put(200,3){\color{black}{$\hat{E}^{t+1}_{b}$}}
        \put(8,3){\color{black}{$\hat{E}^{t}_{b}$}}
        \put(45,167){\color{black}{$E^{t}$}}
        \put(143,169){\color{black}{$\hat{F}_{t\rightarrow t+1}$}}
        \put(145,133.5){\color{black}{$\mathcal{W}$}}
    \end{overpic}
    \caption{An example of using the completed forward flow $\hat{F}_{t\rightarrow t+1}$ to guide the feature backward propagation,
     where $\oplus$ and \sffamily{\textcopyright} \rmfamily denote an addition operation and a concatenation operation, respectively.
     Note that the backward flow will act in the opposite direction.}
    \label{fig:feature_prop}
    \vspace{-5mm}
\end{figure}

\subtitle{Flow-guided feature propagation.}
Suppose $\{{E}^{t} \in \mathbb{R}^{ \frac{H}{4} \times \frac{W}{4} \times C} \mid t=1 \dots T_{l}\}$ are the local temporal neighboring features extracted from the context encoder, where $T_l$ denotes the length of local neighboring frames.
Taking the forward flow $\hat{F}_{t\rightarrow t+1}$ as an example, it assists us in capturing the motion of the corrupted regions from the $t$-th frame to the ($t$+1)-th frame.
Once the pixels in the corrupted regions at the $t$-th content feature is known in the valid area at the ($t$+1)-th feature, we can intuitively exploit this valid information through warping the ($t$+1)-th backward propagation feature $\hat{E}^{t+1}_{b}$ to current time step with the help of the forward flow $\hat{F}_{t\rightarrow t+1}$.
The warped feature can be further merged with current content feature $E^{t}$ and updated through a backward propagation function $\mathcal{P}_{b}(\cdot)$:
\begin{equation}
    \small
    \hat{E}^{t}_{b} = \mathcal{P}_{b}(E^{t}, \mathcal{W}(\hat{E}^{t+1}_{b}, \hat{F}_{t\rightarrow t+1})),
    \label{eq:warpping}
\end{equation}
where $\mathcal{W}(\cdot)$ denotes the spatial warping operation based on optical flow, $\hat{E}^{t}_{b}$ is the backward propagation feature at the $t$-th time step, and the propagation function $\mathcal{P}_{b}(\cdot)$ represents two convolutional layers with a LeakyReLU~\cite{maas2013rectifier} activation.

The warping and merging operations in Eq.~\eqref{eq:warpping} are approximate to the whole propagation process in DFVI and FGVC, but we conduct them in the feature space rather than the image space.
The propagation feature $\hat{E}^{t}_{b}$ is updated step by step as faithful content is gradually involved in the corrupted area for each content feature, which also facilitates the connection across all local neighboring features with flow guidance.
Unlike the hand-crafted pixel-level propagation in flow-based methods, which is very time-consuming and depends heavily on the quality of estimated flow, the feature-level propagation adaptively merges the flow-traced information with larger receptive fields using convolutional layers and can be speeded up by GPUs.

Although the feature-level propagation can be much faster and more effective than FGVC and DFVI, it still needs to face the problem caused by the inaccurate flow estimation results in Eq.~\eqref{eq:flow}, which will bring irrelevant information in the propagation process and further hamper the final performance.
To mitigate this problem, inspired by ~\cite{wang2019edvr,chan2020understanding,chan2021basicvsr,chan2021basicvsr++}, we employ modulated deformable convolution~\cite{zhu2019deformable} to further index and weight the candidate feature points.
As shown in \figref{fig:feature_prop}, we first calculate the weight mask $W_{t\rightarrow t+1}$ and the offsets $\Delta F_{t\rightarrow t+1}$ relative to the estimated optical flow with:
\begin{equation}
    \small
    [W_{t\rightarrow t+1}, \Delta F_{t\rightarrow t+1}] = \mathcal{C}_{b}(E^{t}, \mathcal{W}(\hat{E}^{t+1}_{b}, \hat{F}_{t\rightarrow t+1}), \hat{F}_{t\rightarrow t+1}),
    \label{eq:dcn_warp}
\end{equation}
where $\mathcal{C}_{b}(\cdot)$ denotes multiple cascading convolutional layers.
Both the size of computed weight mask $M_{t\rightarrow t+1}$ and offset $\Delta F_{t\rightarrow t+1}$ are $\frac{H}{4} \times \frac{W}{4} \times K^2 \times G$, where $K$ and $G$ are the kernel size and the group number of deformable convolution, respectively.
We can further generate $K^{2}\times G$ candidate feature points for each spatial location by adding the offset $\Delta F_{t\rightarrow t+1}$ to the completed optical flow $\hat{F}_{t\rightarrow t+1}$.
The relationship between the offset $\Delta F_{t\rightarrow t+1}$ and the completed optical flow $\hat{F}_{t\rightarrow t+1}$ are mutually beneficial.
On the one hand, more flexible sampling locations could well compensate for the inaccurate flow completion.
On the other hand, the completed flow provides promising initial sampling locations, which make it easily find more meaningful content within their surroundings.
Then, we use a deformable convolutional layer to warp the backward feature $\hat{E}^{t+1}_{b}$ instead of optical flow-based warping in Eq.~\eqref{eq:warpping} and further obtain the backward propagation feature $\hat{E}^{t}_{b}$ through:
\begin{equation}
    \small
    \hat{E}^{t}_{b} = \mathcal{P}_{b}(E^{t}, \mathcal{D}_{b}(\hat{E}^{t+1}_{b}, W_{t\rightarrow t+1}, \hat{F}_{t\rightarrow t+1} + \Delta F_{t\rightarrow t+1})),
\end{equation}
where $\mathcal{D}_{b}$ denotes the operation of the deformable convolutional layer.
The weight mask $W_{t\rightarrow t+1}$, whose values are normalized via a sigmoid function, can be applied to each sampling pixel for measuring its validity.

The aforementioned operations are employed bidirectionally following ~\cite{Xu_2019_CVPR,Gao-ECCV-FGVC}, while the forward propagation feature $\hat{E}^{t}_{f}$ can be obtained in the same way but in the opposite direction.
Finally, we use a learnable $1\times1$ sized convolution layer to fuse the forward and backward propagation features adaptively instead of using a pre-defined rule to combine the bidirectional flow traced pixels in ~\cite{Xu_2019_CVPR}.
\begin{equation}
    \small
    \hat{E}^{t} = \mathcal{I}(\hat{E}^{t}_{f}, \hat{E}^{t}_{b}),
    \label{eq:fusion}
\end{equation}
where $\mathcal{I}$ denotes a $1\times 1$ sized convolutional layer.

\subsection{Temporal focal transformer}
\label{sec:focal_transformer}
Only using the information provided by local temporal neighbors is not enough for video inpainting.
As discussed in ~\cite{Gao-ECCV-FGVC}, the corrupted content at local neighbors may appear in the non-local ones.
Thus, the information in the non-local temporal neighbors can be regarded as a promising reference for these missing regions in local neighbors. 
Here we stack multiple temporal focal transformer blocks to effectively combine the information from local and non-local temporal neighbors for performing content hallucination.

Suppose $T_{nl}$ is the number of selected non-local frames.
$\mathbf{E}_{nl} \in \mathbb{R}^{T_{nl} \times \frac{H}{4} \times \frac{W}{4} \times C}$ is the encoded features of all non-local neighbors.
$\mathbf{\hat{E}}_{l} \in \mathbb{R}^{T_{l} \times \frac{H}{4} \times \frac{W}{4} \times C}$ is the local temporal feature through concatenating the results in Eq.~\eqref{eq:fusion} at the temporal dimension.
We use a soft split operation~\cite{Liu_2021_FuseFormer} to perform overlapped patch embedding on the concatenated local and non-local temporal features:
\begin{equation}
    \small
    {Z}^{0} = \operatorname{SS}([\mathbf{\hat{E}}_{l}, \mathbf{E}_{nl}])\in \mathbb{R}^{(T_{l} + T_{nl}) \times M \times N \times C_e} ,
\end{equation}
where  $\operatorname{SS}$ denotes the operation of soft split.
$Z^{0}$ is the embedded token that contains both local and non-local temporal information.
$M \times N$ is the embedded spatial dimension, and $C_e$ is the feature dimension.

Instead of vanilla vision transformer~\cite{dosovitskiy2021an}, which is frequently employed in recent works~\cite{yan2020sttn,Liu_2021_FuseFormer}, we use focal transformer~\cite{yang2021focal} to search from both local and non-local neighbors to fill missing contents.
The reasons are listed as follows:
(1) Compared with performing fine-grained global attention, the computational and memory cost can be effectively reduced through window-based attention~\cite{liu2021Swin,yang2021focal}.
(2) For each token in the missing regions, it is reasonable to perform the fine-grained self-attention only in local regions while the coarse-grained attentions globally because of the local self-similarity of an image.

Since the original focal transformer is unable to process sequence data, we propose a temporal focal transformer that essentially extends the size of focal windows from 2D to 3D.
Specifically, we first split the input token $Z^{n-1}$, where $n \in [1, N]$ and $N$ is the stacking number of focal transformer blocks, into a grid of sub-windows with size $ s_{t} \times s_{h} \times s_{w}$.
The split token $\hat{Z}^{n-1} \in \mathbb{R}^{(\frac{(T_{l} + T_{nl})}{s_{t}} \times \frac{M}{s_{h}} \times \frac{N}{s_{w}} \times C_{e}) \times (s_t \times s_h \times s_w)}$ can be directly used for computing fine-grained local attentions.
To perform global attention at the coarse granularity, a linear embedding layer $f_{p}$ is used to pool the sub-windows spatially via $\hat{Z}^{n-1}_{g} = f_{p}(\hat{Z}^{n-1})\in \mathbb{R}^{(\frac{(T_{l} + T_{nl})}{s_{t}} \times \frac{M}{s_{h}} \times \frac{N}{s_{w}} \times C_{e}) \times s_t}$.
We then calculate the query, key, and value through two linear projection layers $f_{q}$, $f_{kv}$:
\begin{equation}
    \small
    Q^{n} = f_q(\hat{Z}^{n-1}), \quad \{K^{n}_{l}, K^{n}_{g}, V^{n}_{l}, V^{n}_{g}\} =  f_{kv}(\{\hat{Z}^{n-1}, \hat{Z}^{n-1}_{g}\}).
\end{equation}

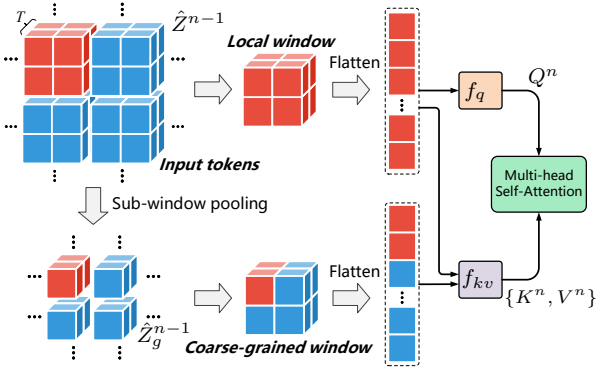
\begin{figure}[t]
    \centering
    \begin{overpic}[width=0.45\textwidth]{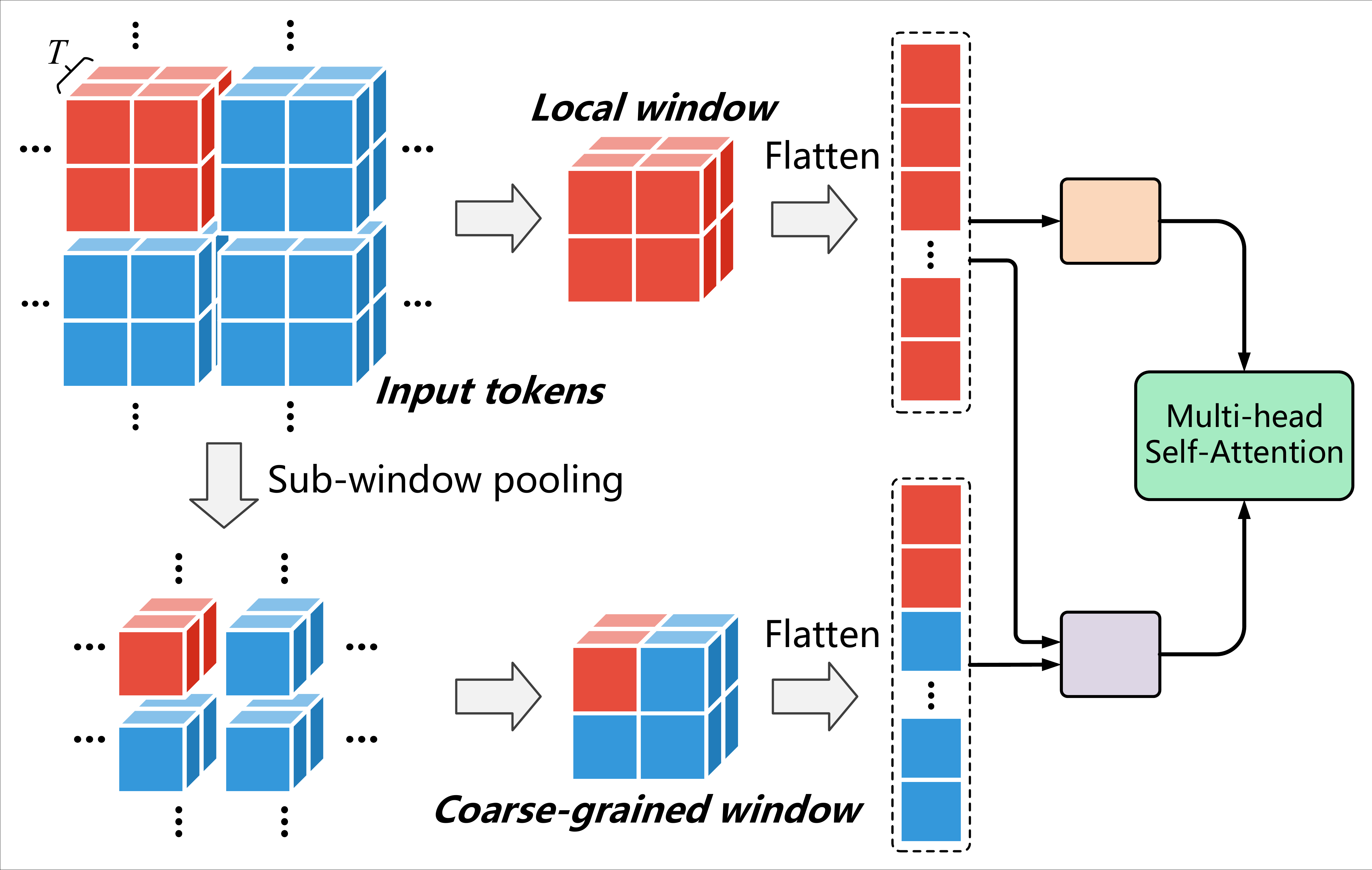}
        \put(176,103){\color{black}{\footnotesize{$f_q$}}}
        \put(175,32){\color{black}{\footnotesize{$f_{kv}$}}}
        \put(65,128){\color{black}{\footnotesize{$\hat{Z}^{n-1}$}}}
        \put(52,10){\color{black}{\footnotesize{$\hat{Z}_{g}^{n-1}$}}}
        \put(200,108){\color{black}{\footnotesize{$Q^{n}$}}}
        \put(191,25){\color{black}{\scriptsize{$\{K^{n}, V^{n}\}$}}}
    \end{overpic}
    \caption{Illustration of temporal focal self-attention. Here we use the window size of $2\times 2\times 2$ as an example. We can see that the keys and values $\{K^{n}, V^{n}\}$ contain both fine-grained local information and coarse-grained global information.}
    \label{fig:focal_attention}
    \vspace{-5mm}
\end{figure}

To calculate attentions with local-global interactions, for the queries inside the $i$-th sub-window $Q^{n}_{i}\in \mathbb{R}^{s_{t} \times s_{h} \times s_{w} \times C_{e}}$, we gather the keys not only from the $i$-th local window $K^{n}_{l,i}\in \mathbb{R}^{s_{t} \times s_{h} \times s_{w} \times C_{e}}$ but also from the $i$-th unfolded coarse-grained window $K^{n}_{g,i}\in \mathbb{R}^{s_{t} \times s_{h} \times s_{w} \times C_{e}}$.
This operation can be processed in parallel.
We concatenate corresponding keys and values respectively by $K^{n}=\{K^{n}_{l}, K^{n}_{g}\}$ and $V^{n}=\{V^{n}_{l}, V^{n}_{g}\}$, and then calculate the focal self-attention for $Q^{l}_{i}$:
\begin{equation}
    \small
    \operatorname{Attention}\left(Q^{n}, K^{n}, V^{n}\right)=\operatorname{Softmax}\left(\frac{Q^{n} (K^{n})^{T}}{\sqrt{C_{e}}}\right) V^{n}.
\end{equation}
Note that the attention function also can work in a multi-head manner.
An example is shown in \figref{fig:focal_attention}.

Finally, the whole process in the $n$-th focal transformer block is formulated as
\begin{equation}
    \small
    Z'^{n} = \operatorname{MFSA}(\operatorname{LN_{1}}(Z^{n-1})) + Z^{n-1},
\end{equation}
\begin{equation} 
    \small
    Z^{n} = \operatorname{F3N}(\operatorname{LN_{2}}(Z'^{n})) + Z'^{n},
\end{equation}
where MFSA and LN denote the multi-head focal self-attention and layer normalization~\cite{ba2016layer}, respectively.
We use F3N~\cite{Liu_2021_FuseFormer} to link the connections across embedded patches.

\subsection{Training objectives}
\label{sec:train_object}
We employ three loss functions to optimize our model.
The first is the reconstruction loss which measures pixel-level differences between synthetic videos $\mathbf{\hat{Y}}$ and the original ones $\mathbf{Y}$ through L1 distance:
\begin{equation}
    \small
    \mathcal{L}_{rec}=\|\mathbf{\hat{Y}}-\mathbf{Y}\|_{1}.
\end{equation}
The second is the adversarial loss which has been proven to be useful for the generation of high-quality and realistic content.
We employ a T-PatchGAN~\cite{chang2019free} based discriminator to make the model focus on both global and local features across all temporal neighbors.
\begin{table*}[t]
    \begin{center}
    \caption{\label{tab:comparison}
        Quantitative comparisons with SOTA video inpainting models on YouTube-VOS~\cite{xu2018youtube} and DAVIS~\cite{perazzi2016benchmark} datasets.
        $\uparrow$ indicates higher is better. $\downarrow$ indicates lower is better.
        ${E_{warp}} ^{*}$ denotes ${E_{warp}}\times 10^{-2}$.
        Each method is evaluated following the procedures in FuseFormer~\cite{Liu_2021_FuseFormer}.
        VINet, DFVI, and FGVC are not end-to-end training methods. Their FLOPs, thus, are not projectable.}
    \vspace{-3mm}
    \renewcommand{\arraystretch}{1.1}
    \renewcommand{\tabcolsep}{2.3mm}
    \scalebox{0.9}{
    \begin{tabular}{l|c|c|c|c||c|c|c|c||c|c}
    \hline
    \cline{1-11}
    & \multicolumn{8}{c||}{Accuracy} & \multicolumn{2}{c}{Efficiency} \\
    \cline{2-11}
    & \multicolumn{4}{c||}{YouTube-VOS} & \multicolumn{4}{c||}{DAVIS} & \multirow{2}{*}{FLOPs}  & Runtime  \\
    \cline{1-9}
    Models & PSNR $\uparrow$ & SSIM $\uparrow$ & VFID $\downarrow$ &  $ {E_{warp}} ^{*}\downarrow$ & PSNR $\uparrow$ & SSIM $\uparrow$ & VFID $\downarrow$ & $ {E_{warp}} ^{*}\downarrow$  & & (s/frame) \\
    \cline{1-11}
    VINet~\cite{kim2019deep}              & 29.20 & 0.9434 & 0.072 & 0.1490 & 28.96 & 0.9411 & 0.199 & 0.1785 & -     & -    \\
    \hline
    DFVI~\cite{Xu_2019_CVPR}              & 29.16 & 0.9429 & 0.066 & 0.1509 & 28.81 & 0.9404 & 0.187 & 0.1608 & -     & 2.56 \\
    \hline
    LGTSM~\cite{chang2019learnable}       & 29.74 & 0.9504 & 0.070 & 0.1859 & 28.57 & 0.9409 & 0.170 & 0.1640 & 1008G & 0.23 \\
    \hline
    CAP~\cite{lee2019cpnet}               & 31.58 & 0.9607 & 0.071 & 0.1470 & 30.28 & 0.9521 & 0.182 & 0.1533 & 861G  & 0.40 \\
    \hline
    FGVC~\cite{Gao-ECCV-FGVC}             & 29.67 & 0.9403 & 0.064 & 0.1022 & 30.80 & 0.9497 & 0.165 & 0.1586 & -     & 2.44 \\
    \hline
    STTN~\cite{yan2020sttn}               & 32.34 & 0.9655 & 0.053 & 0.0907 & 30.67 & 0.9560 & 0.149 & 0.1449 & 1032G  & 0.12 \\
    \hline
    FuseFormer~\cite{Liu_2021_FuseFormer} & 33.29 & 0.9681 & 0.053 & 0.0900 & 32.54 & 0.9700 & 0.138 & 0.1362 & 752G  & 0.20 \\
    \hline
    \hline
    \methodname~(Ours) & \textbf{33.71} & \textbf{0.9700} & \textbf{0.046} & \textbf{0.0864} & \textbf{33.01} & \textbf{0.9721} & \textbf{0.116}  & \textbf{0.1315} & 682G  & 0.16 \\
    \hline
    \cline{1-11}
    \end{tabular}
    }
    \end{center}
    \vspace{-6mm}
\end{table*}
The training objective of this discriminator $D$ is:
\begin{equation}
    \small
    \begin{split}
        \mathcal{L}_{D}=E_{x \sim P_{\mathbf{Y}}(x)}[\mathrm{ReLU}(1-D(x))]+ \\ E_{z \sim P_{\mathbf{\hat{Y}}}(z)}[\mathrm{ReLU}(1+D(z))],
    \end{split}
\end{equation}
For video inpainting generator, the adversarial loss is formulated as:
\begin{equation}
    \small
    \mathcal{L}_{adv}=-E_{z \sim P_{\mathbf{\hat{Y}}}(z)}[D(z)],
\end{equation}
The third loss is the flow consistency loss shown in Eq.~\eqref{eq:loss_flow}.
Training details can be found in supplementary materials.
\begin{figure}
    \centering
    \includegraphics[width=0.4\textwidth]{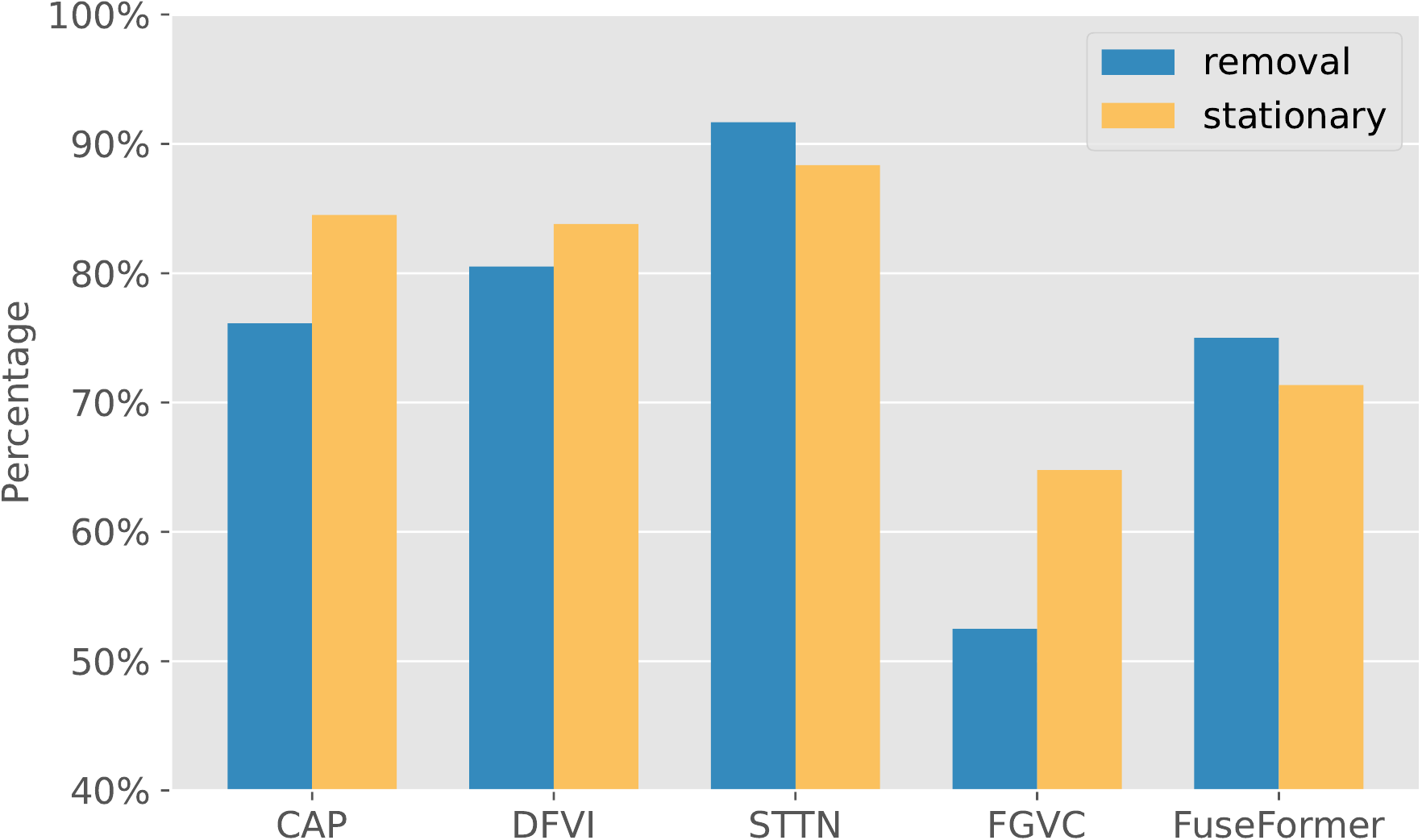}
    \caption{User study results. 
            The vertical axis indicates the percentage of favoring our method compared to other methods.}
    \label{fig:user-study}
    \vspace{-5mm}
\end{figure}

\newcommand{\wh}{.365\columnwidth}
\newcommand{\ww}{.45\columnwidth}
\newcommand{\hcut}{90}
\newcolumntype{L}[1]{>{\raggedright\let\newline\\\arraybackslash\hspace{0pt}}m{#1}}
\newcolumntype{C}[1]{>{\centering\let\newline\\\arraybackslash\hspace{0pt}}m{#1}}
\newcolumntype{R}[1]{>{\raggedleft\let\newline\\\arraybackslash\hspace{0pt}}m{#1}}

\begin{figure*}
    \centering
    \begin{tabular}{C{\wh}C{\wh}C{\wh}C{\wh}C{\wh}}
        Masked Frames & CAP~\cite{lee2019cpnet} & FGVC~\cite{Gao-ECCV-FGVC} & FuseFormer~\cite{Liu_2021_FuseFormer} & \methodname~(Ours) \\
    \end{tabular}
    \includegraphics[width=\textwidth]{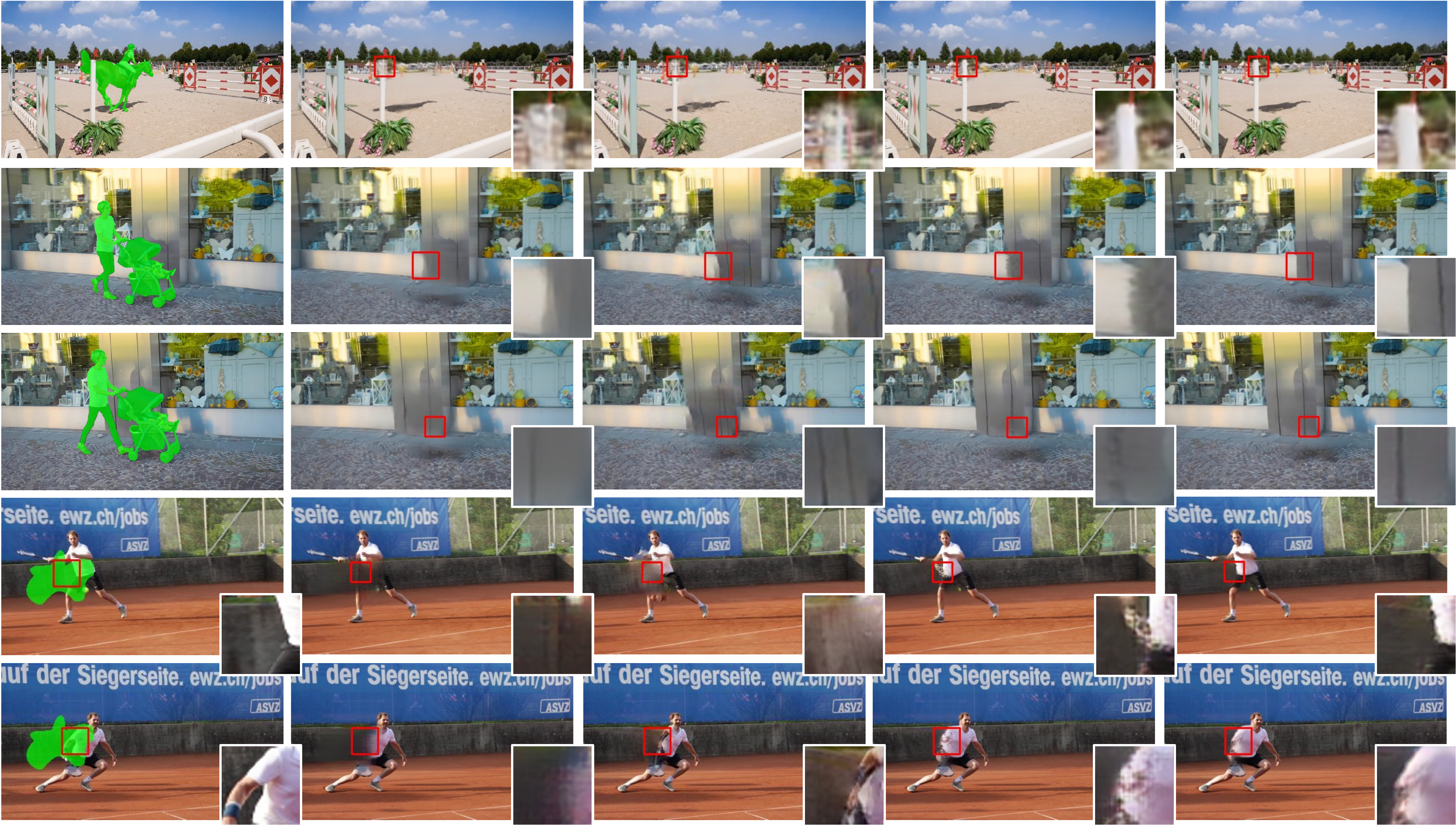}
    \caption{Qualitative results compared with CAP~\cite{lee2019cpnet}, FGVC~\cite{Gao-ECCV-FGVC}, FuseFormer~\cite{Liu_2021_FuseFormer}.}
    \label{fig:comparison}
    \vspace{-5mm}
\end{figure*}

\begin{figure}
    \centering
    \includegraphics[width=0.475\textwidth]{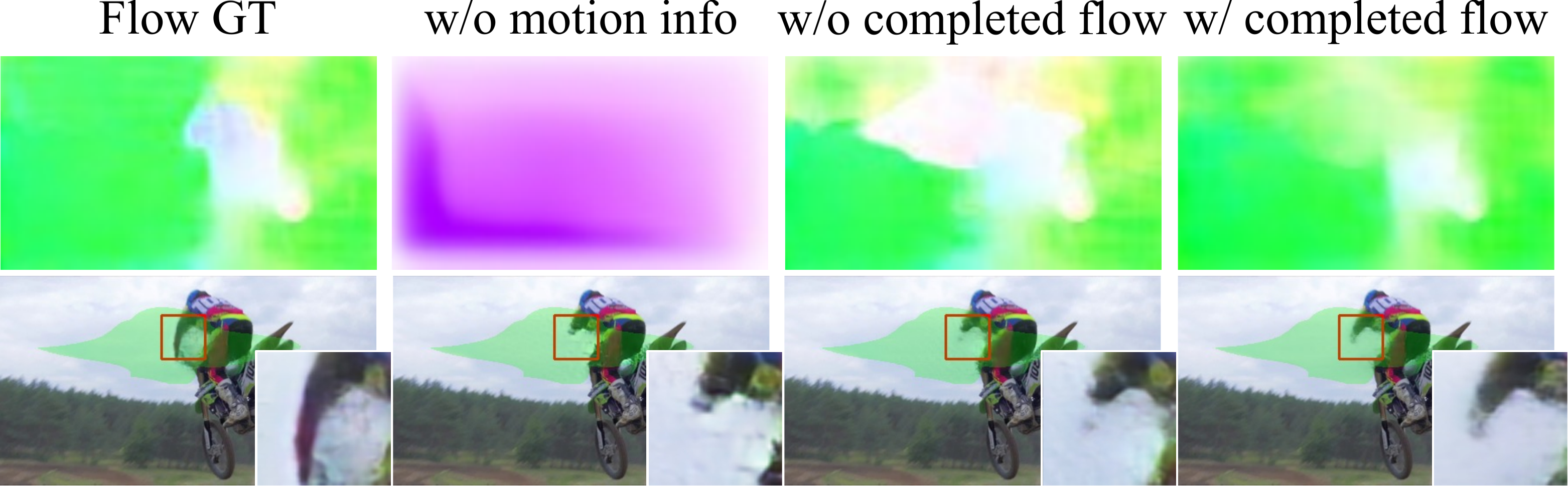}
    \caption{Ablation studies on the flow completion module. 
            The first row shows the results generated from the flow completion modules under different situations.
            The second row visualizes corresponding inpainting frames.}
    \label{fig:flow-abla}
    \vspace{-5mm}
\end{figure}

\section{Experiments}
\subsection{Settings}
\paragraph{Dataset.}
To show the effectiveness of the proposed method, we evaluate it on two popular video object segmentation datasets, \ie, YouTube-VOS~\cite{xu2018youtube} and DAVIS~\cite{perazzi2016benchmark}.
YouTube-VOS, with diverse scenes, consists of 3471, 474, and 508 video clips for training, validation, and test, respectively. 
We follow the original split mode and report the experimental metrics on the test set for YouTube-VOS.
DAVIS is composed of 60 video clips for training and 90 video clips for testing.
Following FuseFormer~\cite{Liu_2021_FuseFormer}, 50 video clips from the test set are used for calculating metrics.
We train our model on the YouTube-VOS dataset and evaluate it on both YouTube-VOS and DAVIS datasets.
As for masks, during training, we generate stationary and object-like masks to simulate video completion and object removal applications following~\cite{kim2019deep,chang2019learnable,lee2019cpnet,yan2020sttn,Liu_2021_FuseFormer}.
For evaluation, stationary masks are used to calculate objective metrics, and object-like masks are adopted for qualitative comparisons because of the lack of references.

\subtitle{Metrics.}
We choose PSNR, SSIM~\cite{wang2004image}, VFID~\cite{NEURIPS2018_d86ea612}, and flow warping error $E_{warp}$~\cite{Lai-ECCV-2018} to evaluate the performance of recent video inpainting methods.
Specifically, PSNR and SSIM are frequently used metrics for distortion-oriented image and video assessment.
VFID measures the perceptual similarity between two input videos and has been adopted in recent video inpainting works~\cite{yan2020sttn,Liu_2021_FuseFormer}. 
Flow warping error $E_{warp}$ is employed to measure the temporal stability.

\subsection{Comparison}
\subtitle{Quantitative results.}
We report quantitative results on YouTube-VOS~\cite{xu2018youtube} and DAVIS~\cite{perazzi2016benchmark} under the stationary masks and compare our method with previous video inpainting methods, including VINet~\cite{kim2019deep}, DFVI~\cite{Xu_2019_CVPR}, LGTSM~\cite{chang2019learnable}, CAP~\cite{lee2019cpnet}, STTN~\cite{yan2020sttn}, FGVC~\cite{Gao-ECCV-FGVC}, and Fuseformer~\cite{Liu_2021_FuseFormer}.
As shown in \tabref{tab:comparison}, our method substantially surpasses all previous SOTA algorithms on all four quantitative metrics.
The superior results demonstrate that our method can generate videos with less distortion (PSNR and SSIM), more visually plausible content (VFID), and better spatial and temporal coherence ($E_{warp}$), which verifies the superiority of the proposed method.

\subtitle{Qualitative results.}
We choose three representative methods, including CAP~\cite{lee2019cpnet}, FGVC~\cite{Gao-ECCV-FGVC}, and Fuseformer~\cite{Liu_2021_FuseFormer}, to conduct visual comparisons.
\figref{fig:comparison} shows both video completion and object removal results.
While the compared methods are hard to recover reasonable details in the masked regions, the proposed method can generate faithful textural and structure information.
This demonstrates the effectiveness of the proposed method.

For further comprehensive comparisons, a user study is conducted on both object removal and video completion applications.
We select five methods including two flow-based methods (\ie, DFVI~\cite{Xu_2019_CVPR} and FGVC~\cite{Gao-ECCV-FGVC}), and three attention-based methods (\ie, CAP~\cite{lee2019cpnet}, STTN~\cite{yan2020sttn}, and Fuseformer~\cite{Liu_2021_FuseFormer}).
We invite 20 participants for the user study totally.
Every volunteer is shown randomly sampled 40 video triplets and asked to select a visually better inpainting video.
Each triplet is composed of one original video, one from our method, and one from a randomly chosen method. 
The user study results are shown in \figref{fig:user-study}.
As we can see, volunteers obviously favor our results over those from almost all methods.
Although such significant preference does not exist in the comparisons with FGVC, the proposed method still receives a majority of votes.
This demonstrates that the proposed method could generate more visually pleasant results than compared methods.

\subtitle{Efficiency comparisons.}
We use FLOPs and inference time to measure the efficiency of each method.
The FLOPs are calculated using the temporal size of 8, and the runtime is measured on a single Titan Xp GPU using DAVIS dataset.
The compared results are shown in \tabref{tab:comparison}.
The proposed method shows comparable running time with transformer-based methods and is nearly $\times 15$ faster than flow-based methods.
Besides, it holds the lowest FLOPs in contrast to all other methods.
This indicates that the proposed method is highly efficient for video inpainting.

\begin{table}
    \begin{center}
        \caption{\label{tab:flow-ablation}
            Ablation studies on the flow completion module.}
        \renewcommand{\arraystretch}{1.2}
        \renewcommand{\tabcolsep}{5mm}
        \scalebox{0.8}{
        \begin{tabular}{c|c|c}
            \Xhline{0.7pt}
            Case                   & PSNR  & SSIM   \\
            \Xhline{0.4pt}
            w/o motion information & 32.08 & 0.9673 \\
            \Xhline{0.4pt}
            w/o completed flow     & 32.23 & 0.9682 \\
            \Xhline{0.4pt}
            w/ completed flow      & 32.35 & 0.9688 \\
            \Xhline{0.4pt}
            Flow GT                & 32.54 & 0.9698 \\
            \Xhline{0.7pt}
        \end{tabular}
        }
        \vspace{-7mm}
    \end{center}
\end{table}

\subsection{Ablations}
We perform three ablation studies on flow completion, feature propagation, and attention mechanism to verify the effectiveness of proposed modules in our framework.
All ablation studies are conducted on the DAVIS dataset.

\subtitle{Study of flow completion module.}
First, we investigate that the importance of motion information for video inpainting.
By only removing the flow consistency loss $\mathcal{L}_{flow}$, our flow completion module no longer provides information about object motions (see \figref{fig:flow-abla}), resulting in a large performance decrease, as shown in \tabref{tab:flow-ablation}.
Second, we study the necessity of completing the optical flow through fixing the pretrained weights in the flow completion module.
With the preliminary knowledge about optical flow, the flow completion module regards the masked regions as occlusion factors and provides initial flow estimation for visible regions (see \figref{fig:flow-abla}).
In contrast to the model without motion information, the performance has an obvious improvement.
However, such model ignores the motion information in the masked regions.
After we complete the flow by training the flow completion module towards minimizing the flow consistency loss, we obtain larger PSNR and SSIM values than before.
As shown in \figref{fig:flow-abla}, the model with completed flows recovers more faithful content about the human arm.
Additionally, in \tabref{tab:flow-ablation} and \figref{fig:flow-abla}, we also show the potential upper bound of our method which estimates the optical flow between uncorrupted frames.

\begin{table}[t]
    \centering
    \caption{\label{tab:prop-ablation}
    Investigation on the feature propagation module. 
    `Flow' indicates the flow-based warping function $\mathcal{W}$ in Eq.~\eqref{eq:dcn_warp}.
    `DCN' denotes modulated deformable convolution~\cite{zhu2019deformable}.}
    \scalebox{0.9}{
        \tabcolsep=0.08cm
        \begin{tabular}{c|c|c|c|c}
            \Xhline{0.7pt}
                 & (a)          & (b)         & (c)          & (d)         \\
            \Xhline{0.4pt}
            Flow & \xmark       & \cmark      & \xmark       & \cmark      \\
            DCN  & \xmark       & \xmark      & \cmark       & \cmark      \\
            \Xhline{0.4pt}
            PSNR & 31.73/0.9653 & 32.15/.9677 & 32.17/0.9676 & 32.35/.9688 \\
            \Xhline{0.7pt}
        \end{tabular}}
    \label{tab:my_label}
    \vspace{-3mm}
\end{table}

\begin{figure}[t]
    \centering
    \includegraphics[width=0.47\textwidth]{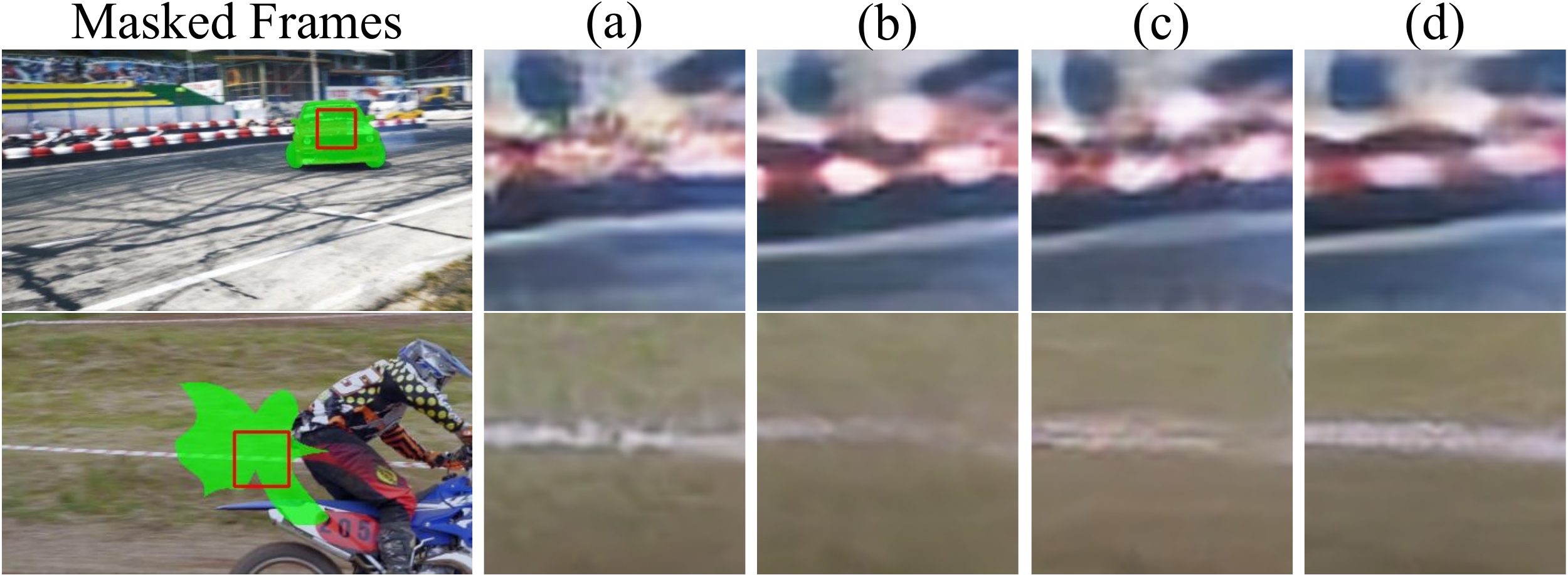}
    \caption{Qualitative results of the ablation studies on the feature propagation module.
    The last four columns correspond to four cases in~\tabref{tab:prop-ablation}.}
    \label{fig:prop-abla}
    \vspace{-6mm}
\end{figure}

\subtitle{Study of feature propagation module.}
After we remove the feature propagation module from the model (case (a) in \tabref{tab:prop-ablation}), the values of quantitative metrics are decreased dramatically. 
From \figref{fig:prop-abla}~(a), we can see that the results generated by this model exist severe artifacts and discontinuous content.
After adding flow-based warping and propagation (see Eq.~\eqref{eq:warpping}) to this model (case (b) in \tabref{tab:prop-ablation}), since we could bring valid pixels from adjacent frames to unseen regions with the assistance of optical flow, the generated content becomes more faithful as shown in \figref{fig:prop-abla}~(b), and the PSNR value is increased by a large margin (0.42dB).
However, it is hard for flow-based warping and propagation to recover the content that cannot be traced by optical flow (the white line in \figref{fig:prop-abla}~(b)).
Besides, for the feature propagation module, which only involves deformable convolution-based warping (case (c) in \tabref{tab:prop-ablation}), the structure details can be more clearly recovered with the help of more learnable offsets, but more artifacts are involved due to the lack of faithful information warped from adjacent frames in contrast to flow-based warping.
By combining deformable convolution with flow guidance (case (d) in \tabref{tab:prop-ablation}), the PSNR and SSIM values can be further improved.
In \figref{fig:prop-abla}~(d), this model achieves the visually best results among all variants while preserving promising structure details. 
This demonstrates the effectiveness of the feature propagation module.

\subtitle{Study of attention mechanism.}
We remove the flow completion and feature propagation modules to purely compare different attention mechanisms, including vanilla global attention (FuseFormer~\cite{Liu_2021_FuseFormer}), local window attention, and focal attention.
As shown in~\tabref{tab:att-ablation}, vanilla global attention achieves the best quantitative performance while suffering from the heavy computation. 
Local attention introduces local windows as Video Swin Transformer~\cite{liu2021video} does.
Although the FLOPs are decreased by 34\%, the attention calculation is limited in the local window, leading to poor performance.
Focal attention shows a good trade-off between performance and computation.
Its PSNR and SSIM values are comparable to FuseFormer, and the computational cost is only increased by 12\% in contrast to the local one.

\begin{table}
    \begin{center}
        \caption{\label{tab:att-ablation}Ablation study on various attention mechanisms.
        FuseFormer~\cite{Liu_2021_FuseFormer} is the current SOTA method that uses vanilla global attention.}
        \renewcommand{\arraystretch}{1.2}
        \renewcommand{\tabcolsep}{1.3mm}
        \scalebox{0.8}{
        \begin{tabular}{c||c|c|c}
            \Xhline{0.7pt}
            Case            & PSNR  & SSIM   & FLOPs \\
            \Xhline{0.4pt}
            FuseFormer      & 31.74 & 0.9662 & 752G  \\
            \Xhline{0.4pt}
            Local attention & 31.57 & 0.9648 & 497G  \\
            \Xhline{0.4pt}
            Focal attention & 31.73 & 0.9653 & 560G  \\
            \Xhline{0.7pt}
        \end{tabular}
        }
    \end{center}
    \vspace{-6mm}
\end{table}

\vspace{-1mm}
\subsection{Limitation}
\begin{figure}[t]
    \centering
    \includegraphics[width=0.47\textwidth]{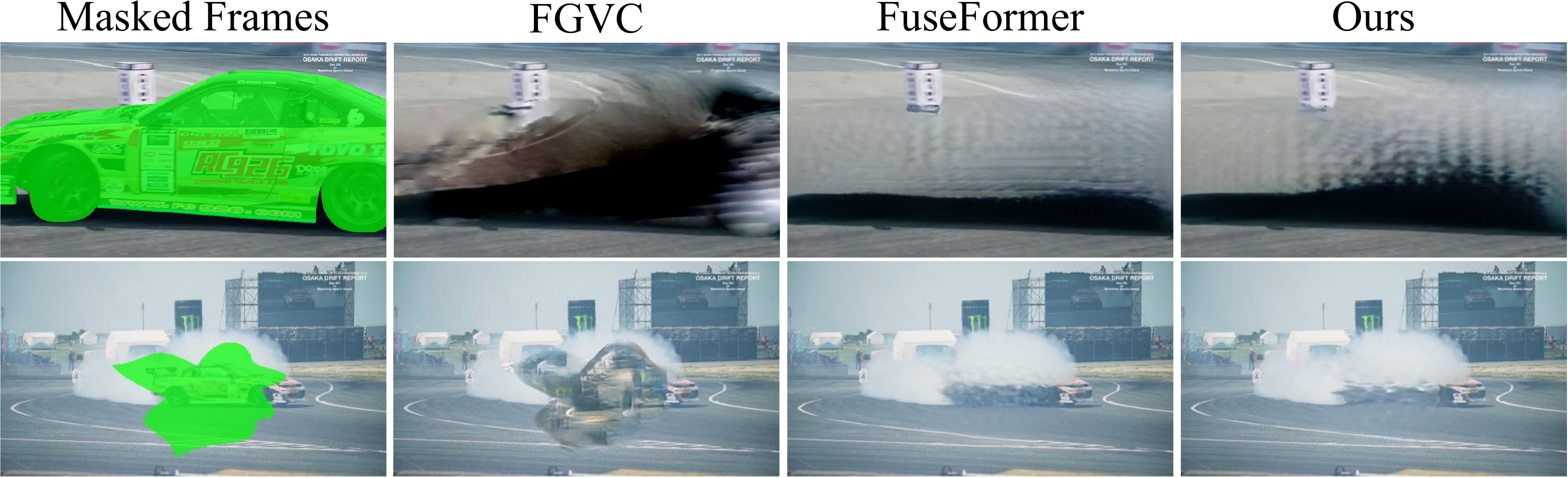}
    \caption{Two failure cases (car drifting).
        Current video inpainting methods fail to deal with large motion or a large number of missing object details and may produce severe artifacts.}
    \label{fig:limitation}
    \vspace{-5mm}
\end{figure}

\vspace{-1mm}
\figref{fig:limitation} shows two failure cases.
When encountering large motion or a large amount of missing object details across frames, our method produces implausible content and many artifacts in masked regions as well as FGVC~\cite{Gao-ECCV-FGVC} and FuseFormer~\cite{Liu_2021_FuseFormer} do.
This demonstrates that these situations are still challenging for video inpainting.
\vspace{-3mm}

\section{Conclusion}
\vspace{-1mm}
We have proposed an end-to-end trainable flow-based model for video inpainting named \methodname.
The elaborately designed three modules (\ie, flow completion, feature propagation, and content hallucination modules) are collaborated together and address many bottlenecks of previous methods.
Experimental results have shown that our method achieves state-of-the-art quantitative and qualitative performance on two benchmark datasets and is efficient in terms of inference time and computational complexity.
We hope it can serve as a strong baseline for future works.

\noindent\textbf{Acknowledgment:}
This work is funded by the National Key Research and Development Program of China (NO. 2018AAA0100400),
NSFC (NO. 61922046),
S\&T innovation project from Chinese Ministry of Education,
and China Postdoctoral Science Foundation (NO.2021M701780).
We also gratefully acknowledge the support of MindSpore, 
CANN, and Ascend AI Processor used for this research.

{\small
\bibliographystyle{ieee_fullname}
\bibliography{ref}
}

\appendix
\section*{Appendix}
\section{Architecture and Training Details}
\paragraph{Architecture.}
In our model, the encoder and the decoder use the same architecture as FuseFormer~\cite{Liu_2021_FuseFormer}.
The channel dim $C$ of the encoder and the decoder is set as 128.
A lightweight model SPyNet~\cite{ranjan2017optical} is employed as our flow completion module for computational efficiency. 
To utilize the learned flow prior in original SPyNet, we use pre-trained weights to initialize this module.
The architecture details of the T-PatchGAN are identical to previous works~\cite{chang2019free,yan2020sttn,Liu_2021_FuseFormer}.
The kernel size $K$ and the group number $G$ of deformable convolution are set as 3 and 16, respectively.
The number of focal transformer blocks $N$ is set as 8 and the embedded dim of tokens $C_{e}$ is set as 512.
The embedded spatial dimension $M \times N$ is $20 \times 36$.
The size of partitioned sub-window $s_{t} \times s_{h} \times s_{w}$ is set to $(T_{l}+T_{nl}) \times 5 \times 9$.
At the end of the content hallucination module, we use a soft composite operator~\cite{Liu_2021_FuseFormer} to composite the embedded tokens to features, which share the same spatial size as the original ones.
\paragraph{Training details.}
For training objectives, the weights of $\mathcal{L}_{rec}$, $\mathcal{L}_{adv}$, and $\mathcal{L}_{flow}$ are 1, $10^{-2}$, and $1$, respectively.
Taking the memory limitations of GPUs into account, we resize all frames from videos into $432 \times 240$ for training, evaluation, and test.
During training, the numbers of local ($T_{l}$) and non-local frames ($T_{nl}$) are 5 and 3, respectively.
Local frames are continuous clips, while non-local frames are randomly sampled from videos for training.
Following STTN~\cite{yan2020sttn} and FuseFormer~\cite{Liu_2021_FuseFormer}, during evaluation and test, we use a sliding window with the size of 10 to get local neighboring frames and uniformly sample the non-local neighboring frames with a sampling rate of 10.
We adopt Adam optimizer with $\beta_1=0$ and $\beta_2=0.99$.
The final model is trained for 500K iterations, and the initial learning rate is set as 0.0001 for all modules and reduced by the factor of 10 at 400K iteration.
In our ablation studies, we train the model for 250K iterations.
We use 8 NVIDIA Tesla V100 GPUs for training and the batch size is set as 8.
Our code is available~\footnote{\url{https://github.com/MCG-NKU/E2FGVI}} for reproducibility.

\section{More Experiments}

\subsection{Completing flows in a offline manner.} 
To verify the effectiveness of online flow completion, we prepare completed flows using the FGVC~\cite{Lai-ECCV-2018} flow completion module in an offline manner. 
We then retrain a model with the FGVC completed flows.
The PSNR value of this model is slightly higher than our end-to-end setting (32.38 vs. 32.35 (dB)).
However, the inference speed is much slower than ours (1.21 vs. 0.16 (s/frame)).

\subsection{Taking a deeper look to flow-guided feature propagation module}
\begin{figure*}[t]
    \centering
    \includegraphics[width=0.9\textwidth]{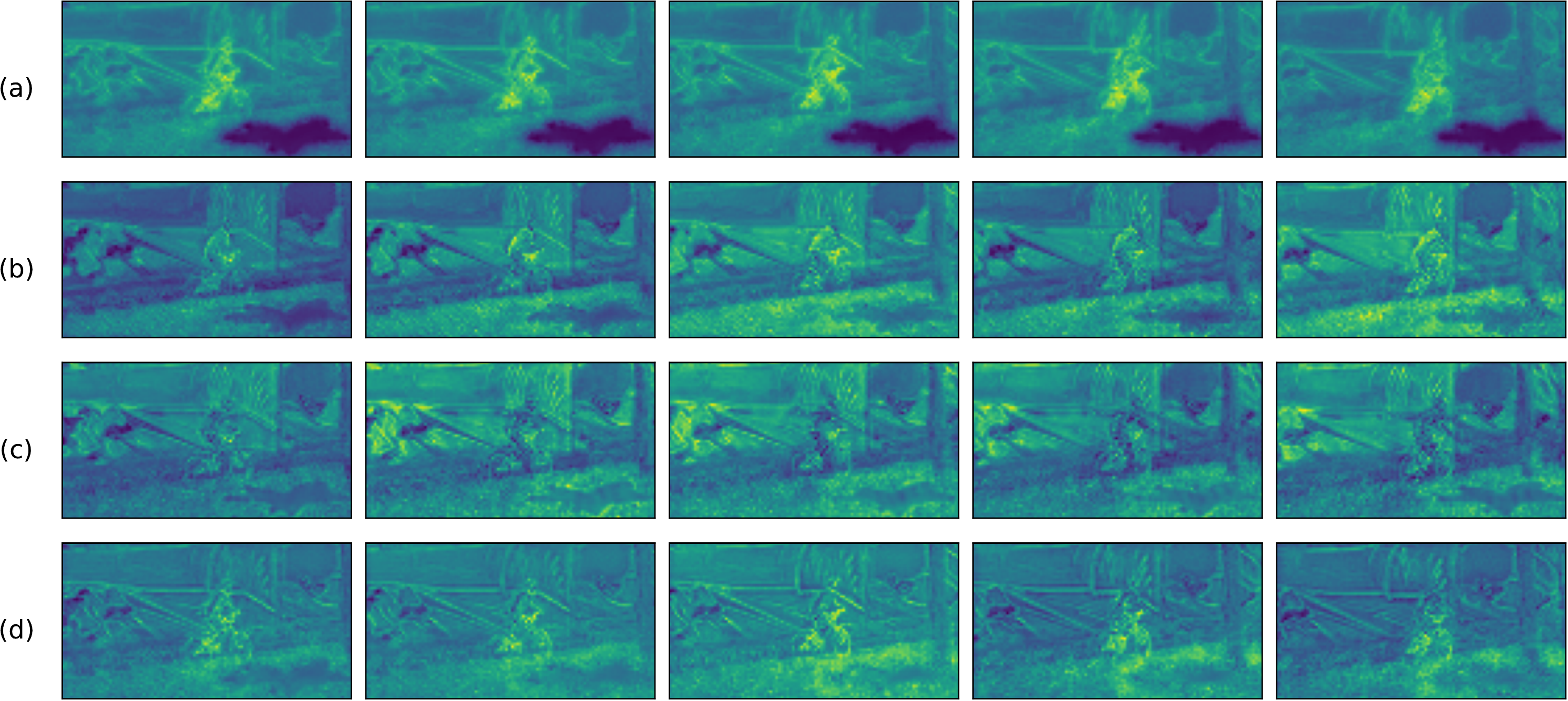}
    \caption{Visualization of the frame-wise average features before feeding into the content hallucination stage under different experimental settings:
    (a) without flow-guided feature propagation, 
    (b) flow-guided feature propagation without deformable convolution (Eq.~3 of the main paper),
    (c) feature propagation without flow guidance,
    and (d) final flow-guided feature propagation module with the assistance of both flow fields and deformable convolution.
    (\textbf{Zoom-in for best view})}
    \label{fig:supp-feat-comp}
\end{figure*}

To further investigate the effectiveness of the feature propagation module, we visualize averaged local neighboring features with the temporal size of 5 before conducting content hallucination in \figref{fig:supp-feat-comp}.
The four cases in \figref{fig:supp-feat-comp} correspond to the four variants in the Tab. 3 of our main paper.
For the model without feature propagation (\figref{fig:supp-feat-comp}(a)), obviously, we can see that corrupted regions from all frames still exist in these features, further restricting the performance of content hallucination.
For the model only using flow-based warping (\figref{fig:supp-feat-comp}(b)) or deformable convolution-based warping (\figref{fig:supp-feat-comp}(c)), corrupted regions are filled with the contents warped from adjacent frames.
And the deformable convolution-based warping can generate smoother content than flow-based one due to more sampling feature points.
However, especially for the last two temporal features (last two columns in \figref{fig:supp-feat-comp}), the regions filled by the model without flow guidance have more distinct boundaries in contrast to flow-based warping, which implies that less faithful content are propagated without motion information.
Through adopting deformable convolution with flow guidance, the final propagation module (\figref{fig:supp-feat-comp}(d)) fills the holes with the most reasonable and natural content among all cases.
This is a promising demonstration of the mutually beneficial relationship between deformable offsets and completed flow fields.

\subsection{Study of the hallucination ability}
To purely evaluate the hallucination ability of our method, we first pre-fill the pixels which can be traced by flow fields~\cite{Gao-ECCV-FGVC}.
The remaining unfill pixels are thus most likely not visible in other video frames.
We then feed the pre-filled videos to an image inpainting model~\cite{yu2018free} and our model, respectively.
Our hallucinated result has a much larger PSNR value than the image inpainting model on DAVIS dataset (31.74 vs. 30.80 (dB)).

\begin{table}[htbp]
    \centering
    \caption{Parameters comparisons. FuseFormer* denotes a larger version of original FuseFormer.}\label{tab:params}
    \resizebox{0.48\textwidth}{!}{\begin{tabular}{|c|c|c|c|}
    \hline
     & FuseFormer~\cite{Liu_2021_FuseFormer} & FuseFormer* & \methodname \\
    \hline
    Params. (M) & 36.6 & 41.6 & 41.8 \\
    \hline
    PSNR/SSIM & 31.74/0.9662 & 31.91/0.9669 & 32.35/0.9688 \\
    \hline
    \end{tabular}}    
 \end{table}

\subsection{Parameter comparison}
We report the parameters in Tab.~\ref{tab:params}.
Although our method consumes $\sim$14\% more parameters than the SOTA method (\ie, FuseFormer~\cite{Liu_2021_FuseFormer}), it achieves a great trade-off between performance and computational complexity among other methods (see \tabref{tab:params}).
For further comparison, we add residual blocks in FuseFormer to achieve similar parameters with ours.
Our method still performs better than the larger version of FuseFormer.

\subsection{More Qualitative Results}
\begin{figure*}[t]
    \includegraphics[width=\textwidth]{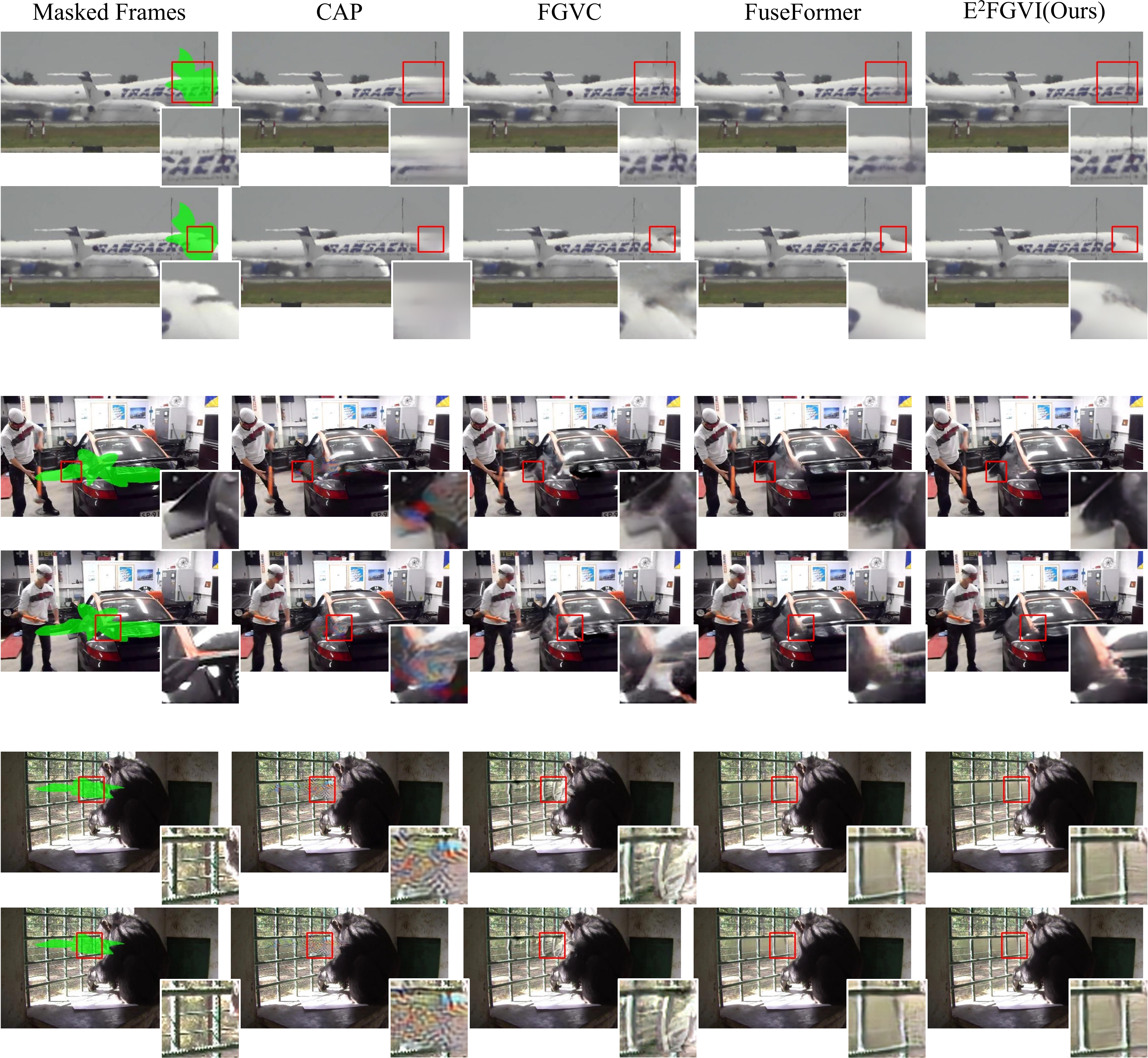}
    \caption{Qualitative video completion results on YouTube-VOS~\cite{xu2018youtube}.}
    \label{fig:sup-comparison-1}
\end{figure*}

\begin{figure*}[t]
    \includegraphics[width=\textwidth]{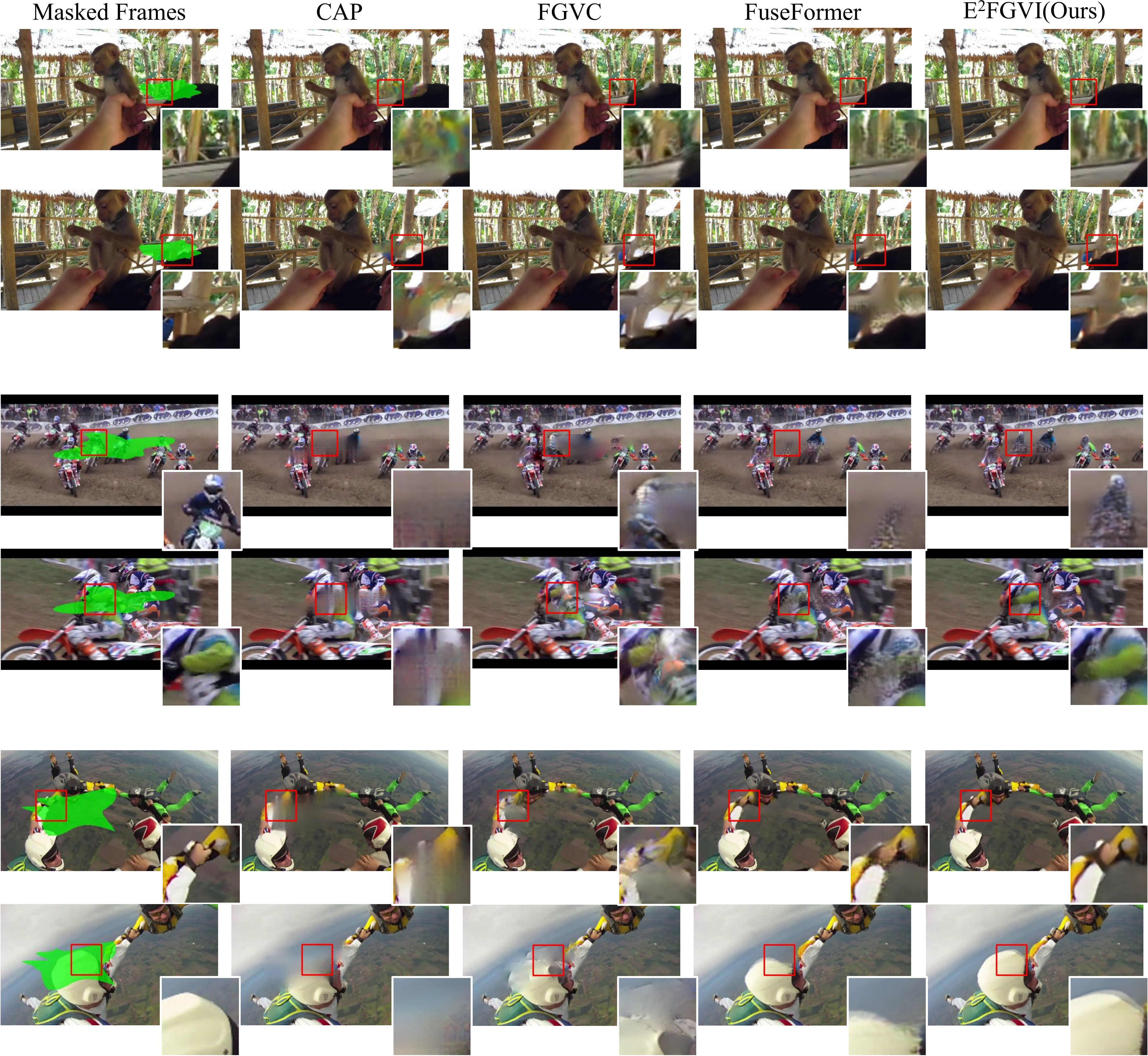}
    \caption{Qualitative video completion results on YouTube-VOS~\cite{xu2018youtube}.}
    \label{fig:sup-comparison-2}
\end{figure*}

\begin{figure*}[t]
    \includegraphics[width=\textwidth]{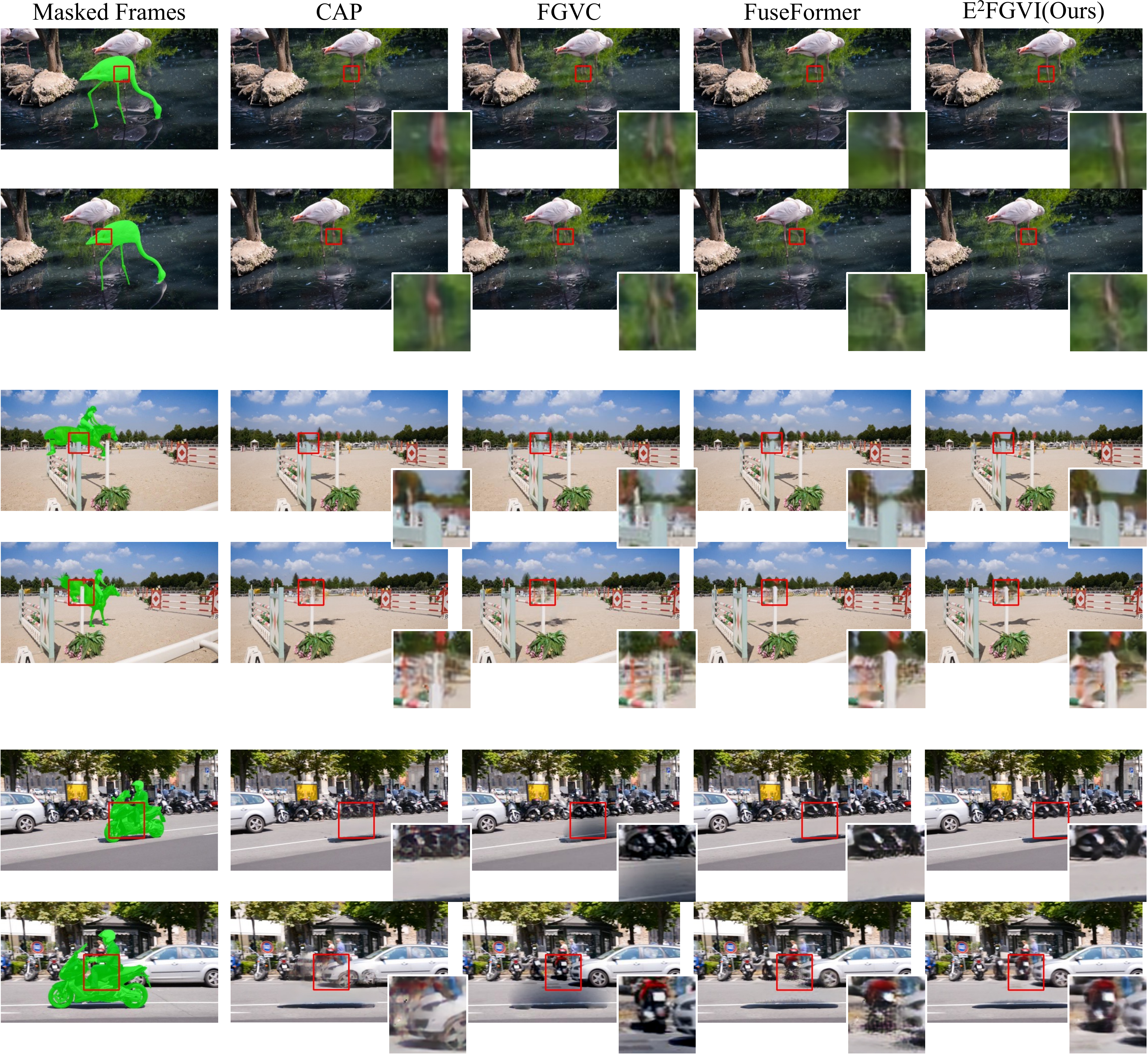}
    \caption{Qualitative object removal results on DAVIS~\cite{perazzi2016benchmark}.}
    \label{fig:sup-comparison-3}
\end{figure*}

\begin{figure*}[t]
    \includegraphics[width=\textwidth]{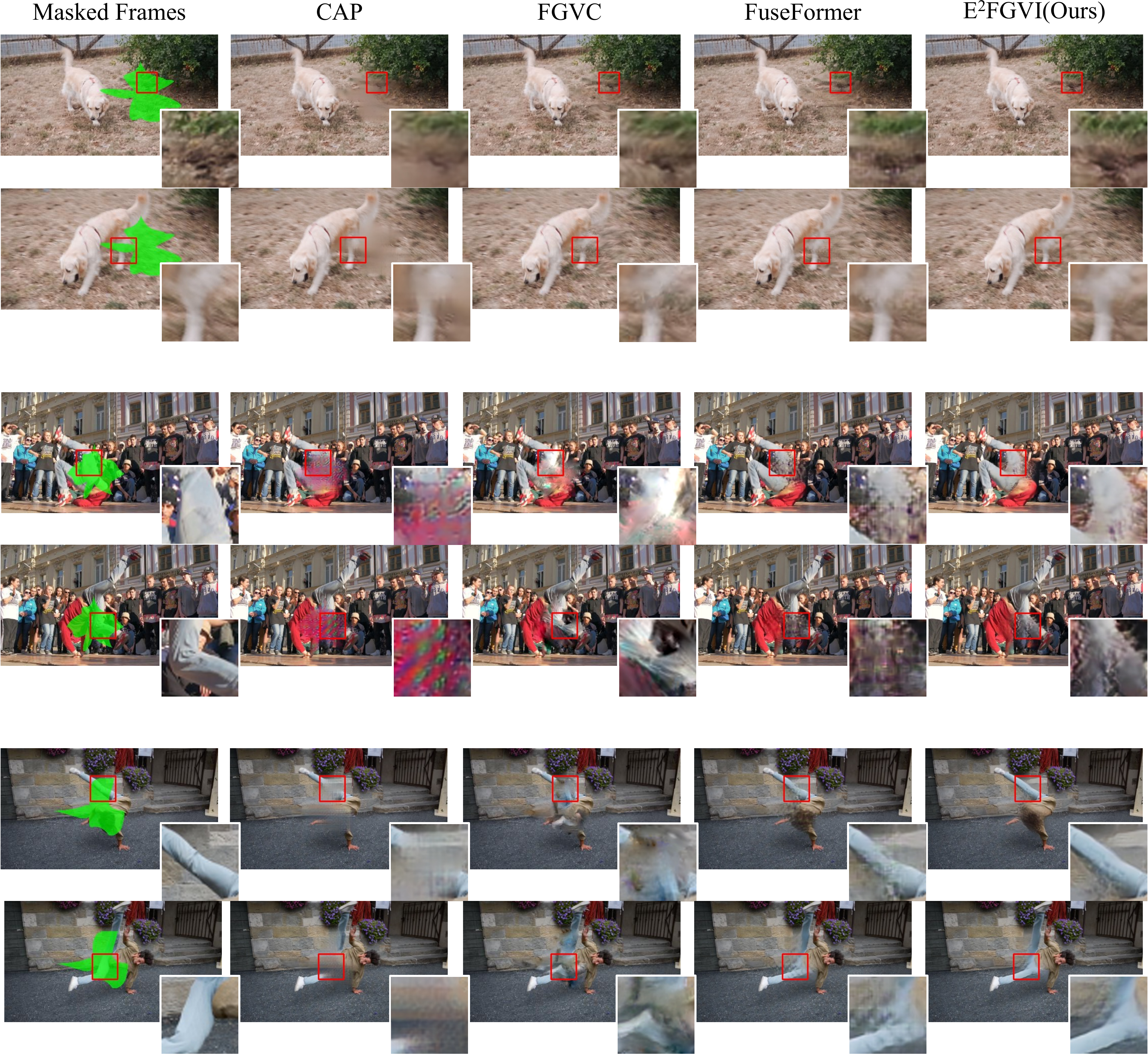}
    \caption{Qualitative video completion results on DAVIS~\cite{perazzi2016benchmark}.}
    \label{fig:sup-comparison-4}
\end{figure*}

In this section, we provide additional visual results on two benchmark datasets, including YouTube-VOS~\cite{xu2018youtube} and DAVIS~\cite{perazzi2016benchmark}, to further show the superiority of the proposed E$^2$FGVI.
The reconstruction results of CAP~\cite{lee2019cpnet}, FGVC~\cite{Gao-ECCV-FGVC}, and FuseFormer~\cite{Liu_2021_FuseFormer} are presented for comparisons. 
As shown in Fig.~\ref{fig:sup-comparison-1}-\ref{fig:sup-comparison-4}, our E$^2$FGVI can generate more faithful textural and structural information and more coherent contents in masked regions than other methods.
\textbf{Our demo is shown in our project page.}

\end{document}